\RequirePackage{amsthm}
\documentclass[sn-mathphys,Numbered]{sn-jnl}

\newcommand\scalemath[2]{\scalebox{#1}{\mbox{\ensuremath{\displaystyle #2}}}}
\usepackage{graphicx}%
\usepackage{multirow}%
\usepackage{amsmath,amssymb,amsfonts}%
\usepackage{mathrsfs}%
\usepackage[title]{appendix}%
\usepackage{xcolor}%
\usepackage{textcomp}%
\usepackage{manyfoot}%
\usepackage{booktabs}%
\usepackage{algorithm}%
\usepackage{algorithmicx}%
\usepackage{algpseudocode}%
\usepackage{listings}%

\usepackage{multicol}%
\usepackage{paralist}%
\usepackage{hyperref}%
\usepackage{comment}%

\theoremstyle{thmstyleone}%
\theoremstyle{thmstyletwo}%

\theoremstyle{thmstylethree}%

\begin{document}

\title[Article Title]{Unveiling Hidden Factors: Explainable AI for Feature Boosting in Speech Emotion Recognition}


\author*[1,2,4]{\fnm{Alaa} \sur{Nfissi}}\email{alaa.nfissi@mail.concordia.ca}

\author[1,4]{\fnm{Wassim} \sur{Bouachir}}\email{wassim.bouachir@teluq.ca}

\author[2]{\fnm{Nizar} \sur{Bouguila}}\email{nizar.bouguila@concordia.ca}

\author[3,4]{\fnm{Brian} \sur{Mishara}}\email{mishara.brian@uqam.ca}

\affil[1]{\orgdiv{Data Science Laboratory}, \orgname{University of Québec (TÉLUQ)}, \orgaddress{\city{Montréal}, \state{Québec}, \country{Canada}}}

\affil[2]{\orgdiv{Concordia Institute for Information Systems Engineering}, \orgname{Concordia University}, \orgaddress{\city{Montréal}, \state{Québec}, \country{Canada}}}

\affil[3]{\orgdiv{Psychology Department}, \orgname{University of Québec at Montréal}, \orgaddress{\city{Montréal}, \state{Québec}, \country{Canada}}}

\affil[4]{\orgname{Centre for Research and Intervention on Suicide, Ethical Issues and End-of-Life Practices}, \orgaddress{\city{Montréal}, \state{Québec}, \country{Canada}}}


\abstract{Speech emotion recognition (SER) has gained significant attention due to its several application fields, such as mental health, education, and human-computer interaction. However, the accuracy of SER systems is hindered by high-dimensional feature sets that may contain irrelevant and redundant information. To overcome this challenge, this study proposes an iterative feature boosting approach for SER that emphasizes feature relevance and explainability to enhance machine learning model performance. Our approach involves meticulous feature selection and analysis to build efficient SER systems. In addressing our main problem through model explainability, we employ a feature evaluation loop with Shapley values to iteratively refine feature sets. This process strikes a balance between model performance and transparency, which enables a comprehensive understanding of the model's predictions. The proposed approach offers several advantages, including the identification and removal of irrelevant and redundant features, leading to a more effective model. Additionally, it promotes explainability, facilitating comprehension of the model's predictions and the identification of crucial features for emotion determination. The effectiveness of the proposed method is validated on the SER benchmarks of the Toronto emotional speech set (TESS), Berlin Database of Emotional Speech (EMO-DB), Ryerson Audio-Visual Database of Emotional Speech and Song (RAVDESS), and Surrey Audio-Visual Expressed Emotion (SAVEE) datasets, outperforming state-of-the-art methods. These results highlight the potential of the proposed technique in developing accurate and explainable SER systems. To the best of our knowledge, this is the first work to incorporate model explainability into an SER framework. The source code of this paper is publicly available via this \href{https://github.com/alaaNfissi/Unveiling-the-Hidden-Factors-Explainable-AI-for-Feature-Boosting-in-Speech-Emotion-Recognition}{https://github.com/alaaNfissi/Unveiling-Hidden-Factors-Explainable-AI-for-Feature-Boosting-in-Speech-Emotion-Recognition}.
}




\maketitle

\section{Introduction}\label{sec1}

The incorporation of emotional aspects into the development of artificial intelligence (AI) has been an area of focus in research for many years. Efforts to incorporate emotion in AI have led to a better understanding of the physiological processes underlying emotions through research in neurophysiological emotional processing \cite{assunccao2022overview}. Emotions are integral to evolutionary and cultural adaptations, influencing human cognition, behavior, and social interactions \cite{totaro2021emotion}. They can manifest as either positive or negative in different scenarios, influenced by the specific context in which they arise. This variation in emotional valence is shaped by the distinct situational elements that trigger these emotional responses \cite{kranzbuhler2020beyond}.

The rise of robotic agents has made it increasingly important to introduce artificial emotionality in social robotics \cite{lim2021social}, as empirical studies have demonstrated improved collaboration and efficiency in task-oriented robotics when artificial agents express emotions to humans \cite{shayganfar2019not} \cite{zhou2020would}. This integration of emotion in AI development holds practical applications, including the ability to monitor user states and mitigate risks through emotion recognition AI. Emotional AI has the potential to enhance psychological interventions in real-time, offering critical support in maintaining homeostatic balance and providing tailored care and rehabilitation for individuals with developmental disorders \cite{gual2022using}.

Psychological research divides emotion modeling into two primary frameworks: discrete categorization, which aligns emotions with a set of fundamental states, and multidimensional approaches, which delve into the qualitative aspects of emotional experiences. Discrete models view emotions as specific archetypal states, whereas multidimensional models analyze emotions in terms of their subjective valence and functional outcomes \cite{cohen2020feel}. The simplicity of emotion representations can make it challenging to distinguish states with similar characteristics \cite{ekman1992argument}. Observable behavioral responses, such as body posture, facial expression, and prosody variation, provide input for models aiming to understand and predict emotional states \cite{totaro2021emotion}.

Speech emotion recognition (SER) involves the application of advanced machine learning techniques to analyze and classify emotions from the various frequencies and features present in speech signals \cite{dougdu2022comparison}. It has pivotal applications across various domains, including enhancing human-computer interactions, advancing affective computing technologies, and contributing significantly to the detection and diagnosis of mental health conditions, as it aids in the accurate interpretation and analysis of emotional states in speech \cite{alsabhan2023human}.

Despite the exploration of numerous machine learning techniques such as support vector machines, hidden Markov models, and deep neural networks for SER, the identification of the most effective feature representations continues to be a significant challenge, given the extensive range of available features and their diverse types \cite{dougdu2022comparison}. A significant obstacle in SER is the training of models on expansive datasets with a wide range of feature representations, where the precise relevance of these features to SER tasks is not always clearly understood. This challenge is further compounded by the limitations of available datasets in terms of size and diversity, hindering the effective training and optimization of machine learning models \cite{abdelhamid2022robust}. Consequently, these predefined feature sets often result in high-dimensional data, making it challenging for models to effectively learn emotion-related patterns.

In the realm of SER, transparency and explainability hold paramount importance, particularly in safety-critical applications. Transparent and understandable SER systems facilitate error identification, correction, and the prevention of potentially dangerous situations \cite{rawal2021recent}. Furthermore, the presence of explainable emotion recognition systems is crucial for producing results that are trustworthy, easily interpretable, and capable of validation. As a result, research in Explainable AI (XAI) strives to enhance transparency and accountability in AI systems, addressing concerns surrounding biases in decision-making processes \cite{samek2019explainable} \cite{mohseni2021multidisciplinary}. Within the context of SER, the exploration of XAI techniques contributes to the development of more transparent and interpretable models in order to foster trust and enhance the overall utility of SER systems.

To address the aforementioned challenges, we propose a comprehensive framework, based on supervised machine learning, that emphasizes feature extraction and selection. Additionally, we integrate an explainability module employing SHapley Additive exPlanations (SHAP) \cite{sundararajan2020many}, to enhance the performance and interpretability of SER systems. It consists of three main modules: 1) a feature boosting module for feature extraction and selection, 2) a supervised classification module for emotion recognition, and 3) an explainability module that explains the model's predictions and evaluates feature contributions using SHAP. The explainability module serves also as a feedback mechanism, continuously refining and boosting the feature set in the first module at each iteration. As far as we know, this is the first study to include model explainability in an SER framework.

Our main contributions can be summarized as follows:
\begin{itemize}
    \item We introduce a novel machine learning approach for SER that prioritizes feature selection through iterative feature boosting, enhancing emotions sparsity via a variation ratio, thus, the model's performance by identifying the most relevant features for emotion recognition.
    \item Our approach incorporates an explainability component that utilizes the SHAP technique to provide transparency and insights into the feature boosting process using a feedback loop. This allows for a better understanding of the contribution of each feature to the final decision.
    \item We conduct an experimental evaluation of our proposed method by comparing it to both human-level performance and state-of-the-art algorithms. This evaluation demonstrates the effectiveness of our approach.
\end{itemize}

\section{Background}\label{sec2}

Previous works on SER have predominantly employed supervised learning methods, which can be categorized into traditional machine learning and deep learning approaches. These methods rely on various handcrafted features and feature selection techniques to classify emotions in speech signals \cite{Iqbal2020MFCCAM, krishnan2021emotion, aggarwal2022two, praseetha2018deep, nfissi2022cnn}.

However, SER is a complex task that comes with several challenges. One of the crucial challenges is effectively representing speech, considering the distinctive patterns that differentiate emotions and the temporal dynamics of emotion expression. Consequently, handling high-dimensional speech data and selecting relevant features become significant obstacles in SER. Extracting a multitude of features without a thoughtful selection process can have detrimental consequences for model performance. This indiscriminate approach may result in overfitting, where the model becomes too specialized in fitting the training data, leading to poor generalization to new data. Overfitting occurs because the excessive features introduce noise and complexity into the model. Conversely, this approach might also lead to subpar performance since it lacks the required sparsity, preventing the model from creating a more generalized and representative understanding of emotions. In essence, a balanced and well-considered feature selection process is crucial to avoid these pitfalls and ensure that the model effectively captures informative and discriminative characteristics of emotions \cite{Iqbal2020MFCCAM, song2020speech, 8785867}.

\subsection{Handcrafted Feature-Based Approaches}

In \cite{Iqbal2020MFCCAM}, authors proposed a method for emotion classification using Mel Frequency Cepstral Coefficients (MFCCs) as features and Support Vector Machines (SVMs) as the classification model. The authors explored the impact of different feature combinations, such as adding pitch and energy features, on classification performance. This study demonstrated the effectiveness of using MFCCs and SVMs for SER, highlighting the importance of feature selection in enhancing classification performance. However, one aspect that could be further improved in the study is the limited exploration of other features. While the authors investigated the impact of adding pitch and energy features to the MFCCs, there may be other relevant features that could contribute to improved classification performance. The study could have benefited from a more comprehensive exploration of feature combinations to ensure that the selected features truly capture the diverse aspects of emotional expression in speech.

Two-way feature extraction approach for SER has been introduced in \cite{aggarwal2022two}. In the first way, authors directly extracted features from audio data using mel-scale related features, which were then dimensionally-reduced using Principal Component Analysis (PCA). The reduced features were fed into a Deep Neural Network (DNN) for classification. The authors observed that PCA helped reduce overfitting and improved DNN training. In the second way, they used the 2D representation of spectrograms for classification, employing the VGG16 CNN model. This approach eliminated the need for feature engineering or selection as it directly utilized the spectrograms as inputs to the model. However, the study does not provide detailed analysis or comparison of the impact of different dimensionalities on classification performance as it briefly mentions the use of feature selection and dimensionality reduction techniques, while the rationale behind the choice of these techniques and especially their impact on classification performance are not adequately discussed.

In \cite{song2020speech} a joint learning framework was proposed for feature selection and emotion classification using the robust discriminative sparse regression (RDSR) approach. This method aimed to select the most discriminative feature subset from the high-dimensional feature set by introducing a feature selection regularization constraint. The authors employed sparse regression to enhance model robustness to outliers and noise. The selected features were then fed into a classification model for emotion prediction. Their experiments demonstrated the superiority of the RDSR approach in terms of classification accuracy and feature selection compared to other state-of-the-art methods. While the study focuses on improving classification accuracy and feature selection, it neglects the aspect of interpretability. It mentions also the use of a feature selection regularization constraint, but it lacks an in-depth explanation of its impact on the feature selection process.
 
To address the sequential nature of speech data, a continuous hidden Markov model (CHMM) was proposed for SER in \cite{8785867}. The model extracted a 33-dimensional feature parameter based on the temporal sequence of speech signals. PCA was then applied to reduce the dimensionality of the feature set. The results showed that the PCA-CHMM model outperformed a standard HMM model that used the entire feature set, demonstrating the effectiveness of dimensionality reduction in improving emotion recognition performance. The study centers on comparing the PCA-CHMM model with a standard HMM model, but it falls short in conducting a comprehensive comparison with contemporary SER models and algorithms. Additionally, the study mentions the use of PCA for dimensionality reduction without offering detailed explanations or justifications for this approach. Moreover, while the study aims to enhance emotion recognition performance, it neglects the aspect of interpretability.

\subsection{Learnable Feature-based Approaches}

Deep learning models have also been applied to SER, leveraging their ability to learn complex features from data. Authors in \cite{nfissi2022cnn} proposed a hybrid end-to-end (E2E) deep learning model combining 1D-CNN and Gated Recurrent Unit (GRU) for feature extraction and classification. The 1D-CNN component extracted spatial features from the input data, while the GRU component captured the time-distributed features and added a time aggregation layer. This model was designed to learn relevant features from raw waveform speech signal and classify emotions in a single E2E process, eliminating the need for handcrafted features. The use of 1D-CNN and GRU components provided a powerful tool for learning complex features in sequential data, which is important for SER as it often depends on the temporal dynamics of speech. While the proposed hybrid E2E model shows promise, the study could benefit from providing more insight into the interpretability of the model's learned features. Understanding the discriminative factors and patterns that contribute to emotion recognition would enhance the trust and applicability of the approach. 

\subsection{Other Approaches}

Authors in \cite{farooq2020impact} integrated traditional handcrafted features with advanced deep learning techniques. This method begins with the extraction of key emotional features from speech data, utilizing conventional handcrafted features known for their effectiveness in capturing the emotional nuances in speech. To augment this, the study incorporates deep CNNs (DCNN), leveraging their capacity for automatic feature extraction from complex datasets. A critical aspect of this methodology is the application of a correlation-based feature selection technique, designed to identify and retain the most discriminative features for SER, thereby enhancing classification accuracy. This approach is further evaluated using a variety of machine learning classifiers, including SVM, random forests, and neural networks. In \cite{pham2021emotion},  the performance of CNNs was investigated on five types of spectral features for classifying an increasing number of emotional categories. They conducted a systematic evaluation of CNN performance on an increasing number of emotions, ranging from binary to eight categories. The authors introduced additional classifications beyond binary or all classes and also proposed a new use of 1D convolution for multiple classes. The study provided insights into CNN performance on an increasing number of emotions and introduced a novel approach using 1D convolution for multi-class classification.
In \cite{ancilin2021improved}, two modifications were proposed to the extraction of MFCCs by using magnitude spectrum instead of energy spectrum and excluding discrete cosine transform while extracting Mel Frequency Magnitude Coefficient. The authors tested these modifications alongside conventional spectral features and evaluated their impact on SER.

Prior studies have emphasized the significance of extracting effective acoustic characteristics to accurately capture different emotional aspects of speech in SER. However, these studies have predominantly relied on pre-defined features without thoroughly examining their relevance for SER or their potential to enhance performance. Certain deep learning-based approaches have attempted to address the issue of feature selection implicitly through 1D convolutions. Meanwhile, other supervised learning methods have sought to tackle the challenge of high dimensionality by applying PCA to compute principal components. In contrast, our proposed approach for supervised SER places a strong emphasis on feature importance and model explainability throughout the entire framework. We aim to explore the relevance of different acoustic features for SER and prioritize the extraction and selection of the most pertinent features. Additionally, we introduce an interpretable machine learning model that provides insights into how the model makes predictions. By explicitly addressing the feature selection challenge in SER and highlighting the importance of explainability in machine learning models, our approach offers a novel perspective in this domain.

\section{Proposed Framework}\label{sec3}

\subsection{Motivation}
\label{motiv}
In the field of SER, the effectiveness of any predictive model is intrinsically tied to the delicate balance and quality of its feature set. Our research is driven by critical questions that confront core challenges in SER: How can we optimize the selection of features to enhance the model's sensitivity to emotional nuances in speech? What strategies can we employ to avoid the pitfalls of overfitting and underfitting, ensuring our model's robustness and adaptability to new, unseen data? Furthermore, we investigate the role of feature diversity in capturing the broad spectrum of human emotions conveyed through speech. This exploration is not just about achieving high accuracy in emotion classification but also about understanding the underlying patterns that govern emotional expression in human speech. To this end, we delve into the intricacies of voice signal characteristics, seeking to identify and utilize those features that are most indicative of emotional states. Another dimension of our research addresses the transparency of SER models. The interpretability of our model's predictions is paramount, providing insights into the decision-making process and helping to identify the most influential features. This approach aligns with the growing emphasis on explainable AI, where understanding the 'why' behind a model's predictions is as vital as the predictions themselves.

\subsection{Overview}
\label{over-v}

\begin{figure*}[t]
\centering
\scalemath{1}{
\centerline{\includegraphics[width=1\textwidth]{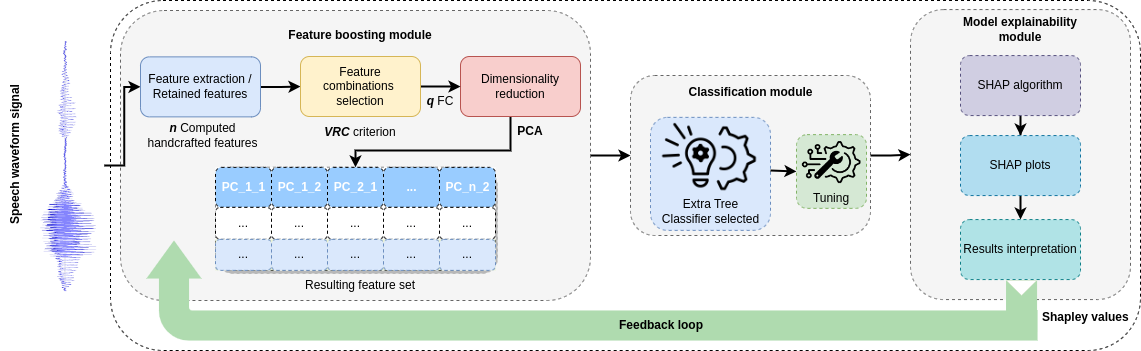}}
}
\caption{The proposed method with its main modules: a) Fature boosting module, b) Classification module, c) Model explainability module. FC stands for Feature Combination}
\label{fig1}
\end{figure*}

Our approach comprises three interrelated components that synergistically improve SER, as illustrated in Fig. \ref{fig1}. The first component is a feature boosting module, which extracts an initial feature set with potential relevance for emotion recognition. This module identifies the most effective feature combinations that best separate emotion classes through a sparsity criterion and then reduces the dimensionality of the resulting feature set while retaining the majority of the relevant information. The second component formulates the SER task as a supervised classification problem, utilizing a classification module to construct a model that can categorize speech samples based on the extracted features. Finally, the third component incorporates an explainability module that analyzes the classification decisions. By leveraging Shapley values, this module gains insights into the feature boosting process and evaluates the significance of features in the decision-making process. One significant advantage of utilizing SHAP over other approaches such as LIME \cite{ribeiro2016should} for model explanations is the type of explanation they offer. While LIME is model-agnostic, which primarily offers local explanations tailored to individual predictions, SHAP is a model-specific method offering global explanations that illuminate the model's behavior across the entire dataset. This holistic perspective is invaluable for understanding how the model considers various features and interactions, identifying dataset-level feature importance, and revealing broader patterns and tendencies. When prioritizing transparency, interpretability, and the ability to assess a model's overall performance, SHAP's provision of global explanations makes it a preferable choice.

Furthermore, our approach incorporates an iterative feedback mechanism that facilitates information exchange between the third component, the explainability module, and the first component, the feature boosting module. This iterative process allows for the continuous refinement and enhancement of the retained features. By analyzing the classification decisions and evaluating the importance of features through the explainability module, we can identify the most relevant features and discard the less relevant ones. This iterative feedback loop ensures that the feature set becomes increasingly optimized over time. Our integrative approach optimizes feature selection, enables supervised classification, and provides interpretability by analyzing classification decisions, thereby improving the overall performance of our SER approach.

In addressing these questions, our framework introduces a comprehensive methodological approach that encompasses three integral components: Feature Boosting Module, Classification Module, and Explainability Module. Each module is meticulously designed to tackle specific aspects of the SER challenge. The Feature Boosting Module focuses on extracting and refining a feature set that captures the essence of emotional states in speech. The Classification Module then takes these features and applies them in a supervised learning context to categorize emotional expressions accurately. Finally, the Explainability Module provides a window into the model's inner workings, offering insights into how and why certain features play a pivotal role in emotion recognition. This integrative approach not only aims to enhance the accuracy and efficiency of emotion classification but also ensures that the process is transparent and understandable.

\subsection{Method}
\subsubsection{Feature boosting module}
In the first step, we calculate a preliminary feature set representation, including pitch, energy, and rhythm-related characteristics, which we assume are meaningful for the SER task. We also calculate statistical characteristics such as the mean, median, standard deviation, minimum, and maximum based on previous works \cite{kacur2021speech} \cite{koduru2020feature}, resulting in a set of $n$ initial features.

In order to improve both the performance and interpretability of our technique, we employ an initial feature selection process. This process starts by selecting a set of $m$ distinct feature combinations of $p$ features from the initial feature set. Then, to determine the most informative and discriminative feature combinations that enhance the separation between emotion categories, we introduce the Variance Ratio Criterion (VRC) as a sparsity criterion. The VRC is designed for datasets with multiple emotion categories, denoted by $E$, and utilizes a dataset $D$ containing $N$ data samples represented as $d_i$. This criterion evaluates the similarity of a speech sample to its own emotion category (cohesion) compared to other emotion categories (separation), as formulated in eq. \ref{eq_vrc1}:

\begin{equation}
    VRC = \frac{\left[\sum_{e=1}^E n_e\left\|c_e-c\right\|^2\right] / (E-1)}{\left[\sum_{e=1}^E \sum_{i=1}^{n_e}\left\|d_i-c_e\right\|^2\right] / (N-E)}
    \label{eq_vrc1}
\end{equation}

\noindent where, $n_e$ represents the number of data points in the $e^{th}$ emotion category, and $c_e$ is the centroid of the data points in that category. The global centroid of all data points is denoted as $c$. The VRC provides a measure of how well-separated and dense the emotion classes are, with higher values indicating better separation.

To determine the significance of each feature combination, we compute the difference between the VRC of the feature combination ($VRC_{fc_i}$) and the VRC of the previous overall feature set ($VRC_{all}$). This difference, denoted as $\sigma_{i}$ in eq. \ref{eq_vrc2}, quantifies the improvement in separation achieved by the specific $i^{th}$ feature combination.

\begin{equation}
    \sigma_{i} = VRC_{fc_i} - VRC_{all}
    \label{eq_vrc2}
\end{equation}

To select the relevant feature combinations, we compare each $\sigma_{i}$ value to a threshold $\alpha$ as calculated in eq. \ref{eq_vrc3}. Specifically, we rank the $m$ combinations in decreasing order of their $\sigma_{i}$ values and retain the top $q$ combinations that satisfy the condition $\sigma_{i} \geq \alpha$, with

\begin{equation}
    \alpha = \frac{\sum_{j=1}^{m_p} \sigma_{j}}{m_p} + \epsilon
    \label{eq_vrc3}
\end{equation}

\noindent where $m_p$ is the number of combinations with $\sigma_{j} \geq 0$, and $\epsilon$ is a parameter we introduce to control the retention of feature combinations based on their explained variance ratio. By using the VRC and the significance measure $\sigma_{i}$, we can identify the feature combinations that contribute the most to the separation and the density of emotion categories, leading to improved emotion recognition performance.

We then apply PCA to reduce the dimensionality of each selected feature combination and eliminate noise. This is achieved by transforming the feature combinations using an eigenvector matrix ($A_i$) and a corresponding eigenvalue vector ($\lambda_i$). The calculation of $A_i$ and $\lambda_i$ for each combination $i$ follows the equations specified as \ref{eq1} and \ref{eq2}. By applying this transformation, we effectively reduce the dimensionality of the feature space while preserving the most informative characteristics.

\vspace{-0.5cm}
\begin{multicols}{2}
\begin{equation}
A_i =
\scalemath{0.72}{
\begin{bmatrix}
a_{i11} & a_{i12} & .. & a_{i1r_i} \\
.. & .. & .. & .. \\
.. & .. & .. & .. \\
a_{ip1} & .. & .. & a_{ipr_i}
\end{bmatrix}
}
\label{eq1}
\end{equation}

\begin{equation}
\lambda_i =
\scalemath{0.72}{
\begin{bmatrix}
\lambda_{i1} \\
.. \\
.. \\
\lambda_{ip}
\end{bmatrix}
}
\label{eq2}
\end{equation}
\end{multicols}
\noindent where, $a_{ijk}$ is the $k^{th}$ element of the $j^{th}$ eigenvector of the $i^{th}$ feature combination and $\lambda_{ij}$ is the eigenvalue associated with the $j^{th}$ principal component of the $i^{th}$ feature combination. Each column of the matrix $A_i$ represents the $j^{th}$ principal component ($PC_{ij}$) of the $i^{th}$ feature combination, capturing specific data information and determining the dimension ($r_i$) of the reduced subset, as shown in eq. \ref{eq4}:

\begin{equation}
PC_{ij} = (a_{i1j})X_{i1} + \dots + (a_{ipj})X_{ip} \label{eq4}
\end{equation}
where, $X_{ik}$ is the $k^{th}$ feature of the $i^{th}$ combination. Therefore, we can determine which features contribute the most to each principal component, which helps us identify the best combination of features representing information in our dataset.

We evaluate the percentage of variance explained by the $j^{th}$ principal component of the $i^{th}$ combination ($PC_{ij}$) using eq. \ref{eq3}:

\begin{equation}
EV_{ij} = \frac{\lambda_{ij}}{\sum_{j=1}^{p}\lambda_{ij}} \times 100 \label{eq3}
\end{equation}
\noindent where, $EV_{ij}$ represents the percentage of variance explained by the $j^{th}$ principal component of the $i^{th}$ combination, and $\lambda_{ij}$ is the eigenvalue and amount of variance explained by $PC_{ij}$.
Finally, we construct a new feature set consisting of the principal components ($PC_{ij}$) of the selected feature combinations. 

This module encompasses several steps, including computing a preliminary feature set, selecting feature combinations, ranking them based on improvement in separation, and constructing a new feature set with selected principal components (refer to algorithm \ref{algo1}-\ref{fb}). It effectively reduces dimensionality while preserving informative features for classification. The process is iteratively enhanced through the feedback loop of the explainability module, as explained in section \ref{xai-mod}.

\subsubsection{Classification module}
\label{class-mod}

In the proposed SER framework, the classification module is integral, where extracted features are used to discern emotions in speech. This module encompasses a diverse set of $M$ candidate classification models, as detailed in section \ref{exp-setp}, each chosen for its proficiency in handling complex emotional data from speech and selected for their adeptness in previous studies processing high-dimensional feature spaces inherent in SER tasks.

The methodology involves inputting the optimized feature set, refined by the feature boosting module, into these models. This step is crucial as it ensures the models operate with features most indicative of emotional states. The training phase for each model involves learning the correlations between these features and the emotional labels within the training dataset portion. This is followed by a validation phase on a separate unseen dataset segment, a crucial step to assess the models' generalizability and accuracy.

Performance assessment is conducted through various metrics as detailed in \ref{eval-mtrc}, providing a comprehensive evaluation of each model's classification efficacy. Subsequent to performance evaluation, the module engages in a detailed optimization process. This phase involves hyperparameter tuning and model refinements, utilizing grid search and cross-validation techniques to enhance the predictive performance of each model.

\begin{algorithm}[H]
\caption{Speech Emotion Recognition (SER) Algorithm}
\label{algo1}

\textbf{Input:}
\begin{itemize}
    \item Speech dataset with labeled emotional states
    \item Selected feature combination size $(m)$
    \item Number of selected principal components per combination $(p)$
    \item Threshold parameter $(\epsilon)$
    \item Number of candidate classification models $(M)$
    \item Convergence criteria
\end{itemize}
\textbf{Output:}
\begin{itemize}
    \item Trained SER model
    \item Retained features
\end{itemize}
\textbf{Procedure:}
\begin{enumerate}[\bfseries 1)]
    \item Preprocess the speech dataset (e.g., filtering, normalization).
    \item Calculate the preliminary feature set with pitch, energy, rhythm, and statistical characteristics.
    \item While not converged:
    \label{iter}
    \begin{enumerate}[\bfseries a)]
        \item Apply feature boosting:
        \label{fb}
        \begin{itemize}
            \item Select $m$ optimal combinations of $p$ features.
            \item Compute $\sigma_i$ and select top $q$ combinations.
            \item Apply PCA to reduce dimensionality and eliminate noise.
            \item Construct a new feature set with the selected principal components.
        \end{itemize}
        \item Train and evaluate $M$ candidate classification models:
        \begin{itemize}
            \item Split the dataset into training and validation sets.
            \item For $i = 1$ to $M$:
            \begin{itemize}
                \item Train the $i^{th}$ model on a new feature set.
                \item Evaluate performance on the validation set.
            \end{itemize}
            \item Select the best-performing model.
        \end{itemize}
        \item Analyze classification decisions using the explainability module:
        \begin{itemize}
            \item Calculate Shapley values to determine the contribution of each principal component.
            \item Identify the most important principal components and feature combinations.
            \item Identify the most important features from the previous feature set.
        \end{itemize}
        \item Update the previous feature set by removing less important features.
        \item Check for convergence by comparing with previous iterations.
    \end{enumerate}
    \item Train optimal SER model on the entire dataset using retained features.
    \item Output trained SER model and retained features.
\end{enumerate}
\textbf{End}.
\end{algorithm}

After optimizing the models, a comparative analysis is conducted to select the most effective model. This selection is based on various criteria, including not only the accuracy but also the model's adaptability to diverse speech patterns. The finalized model is then integrated with the explainability module, employing Shapley values to illuminate the decision-making processes within the model. This integration is vital, as it provides insights into feature contributions and informs further refinements.

The classification module, thus, is not merely a collection of machine learning algorithms; it represents a structured process. This module embodies a fusion of classification methodologies and domain-specific optimizations, ensuring that the SER system achieves a good performance. It is worth noting that this approach is not limited to SER but can be adapted to various classification tasks, allowing for the evaluation of alternative candidate models in different domains. It provides a valuable means to select the most appropriate model that maximizes performance with the boosted features, contributing to more accurate and reliable classification outcomes.

\subsubsection{Explainability module}
\label{xai-mod}
The explainability module incorporates XAI capabilities into the SER system. This helps not only to create a system that is transparent and understandable in terms of prediction and decision-making but also to provide the feature boosting module with the necessary insights for enhanced data representation. To achieve this, we use the Shapley explanation values to explain the model's predictions. Shapley values allow us to understand the contribution of each $PC_{ij}$ in the resulting feature set to a model's prediction. By using these values, we can identify which combination's principal components are most important for SER.

The interpretability of a model's predictions is crucial for both validation and practical application. This is where the integration of Shapley's values becomes indispensable. Shapley values offer a robust, mathematically grounded method to quantify the contribution of each feature within a complex, multivariate SER model. Each feature in a speech sample contributes to the overall emotional classification, but the extent and nature of this contribution can be elusive in high-dimensional spaces typical of SER models \cite{tharwat2020classification}.

Shapley values address this by distributing the 'payout' (i.e., the prediction output) among the features, based on their marginal contribution to the prediction across all possible combinations of features. This method aligns with the cooperative game theory, where each feature is considered a 'player' in the game of classification \cite{lundberg2017unified}. By employing this approach, we can dissect the model’s decision-making process, revealing how each feature influences specific emotion classifications - whether a certain tone of voice is pivotal in identifying sadness, or if a particular speech rhythm is key to detecting excitement.

In SER, this granular insight is invaluable. It allows for a nuanced understanding of the feature interactions within complex emotional spectra, guiding the refinement of feature engineering and selection processes \cite{rudin2019stop}. Moreover, in scenarios where SER models need to be transparent and their decisions interpretable - such as in user-centric applications or clinical settings - Shapley values provide a scientifically rigorous explanation. They enable us to present a clear, quantifiable rationale behind each prediction, improving the credibility and utility of SER systems \cite{murdoch2019definitions}.

The contribution of each $PC_{ij}$ in the resulting feature set is denoted as $\phi_{PC_{ij}}$ and formulated in eq. \ref{eq6}:
\begin{equation}
    \phi_{PC_{ij}} = \sum_{S\subseteq {11,\dots,qJ_{i}}\backslash{ij}}\frac{1}{|S|}\sum_{s\subseteq S}(-1)^{|s|+1}f_{ij}(z_{s\cup{ij}})-f(z_{s})
    \label{eq6}
\end{equation}
\noindent where, $q$ is the number of combinations, $S$ is a subset of principal component indices, $J_{i}$ is the number of principal components used from each combination $i$ with $J_{i}\geq1$, $z$ is the input vector, and $f(z)$ is the output of the classification model for input $z$. The first summation term $(f_{ij}(z_{s\cup{ij}}))$ computes the expected output of the model when $PC_{ij}$ is included in the subset, while the second term $(f(z_{s}))$ computes the expected output of the model when $PC_{ij}$ is excluded from the subset of principal components $S$. The difference between these two terms represents the contribution of $PC_{ij}$ to the output, which is the marginal contribution of $PC_{ij}$ to coalition $S$. This value is used to conduct a feature importance analysis, which provides insight into how the model works and what factors are most important.

The explainability module tackles one of the most pressing questions in SER: How does the model arrive at its conclusions? By implementing Shapley values, this module demystifies the model's decision-making process, elucidating the significance of each feature in the prediction. This not only enhances our understanding of the model's functionality but also informs ongoing refinement efforts, creating a feedback loop that continuously improves the feature set's effectiveness.

\subsubsection{Feedback mechanism}

The feedback mechanism within our method allows us to refine the feature selection process and enhance the performance and interpretability of the SER system. It operates iteratively to identify the most relevant principal components that effectively capture essential information for emotion recognition. The contribution of features at iteration $t$ to each relevant principal component is carefully evaluated to determine their significance in both the principal components and the overall classification decision process. As a result, features at iteration $t$ that do not significantly contribute to the system's performance are eliminated in the following iteration $t+1$, which is done progressively at each iteration until convergence is achieved. This iterative approach enables us to continuously improve the accuracy and interpretability of the SER system by selectively retaining the most informative features and discarding less important ones. The detailed steps of this iterative feedback mechanism can be found in section \ref{iter} of algorithm \ref{algo1}.

\section{Experiments and results}
\subsection{Experimental design}
\subsubsection{Datasets}

In the development of SER systems, the selection of speech datasets is crucial, encompassing three primary categories: actor-based, induced, and natural emotional datasets. The fidelity of SER models largely hinges on the authenticity of these datasets, with natural ones offering genuine emotional representations. However, challenges such as data accessibility and legal concerns often accompany the use of natural datasets, underscoring the need for careful dataset selection in SER model development \cite{dougdu2022comparison} \cite{kumaran2021fusion}.

Emotional speech datasets vary across several properties such as language, database type, number of emotions, data type, number of speakers, number of samples, number of utterances, and dataset aim \cite{singh2022systematic}. Among all the datasets, actor-based ones account for the majority of research datasets. The most commonly used language is English, followed by Chinese and German. These datasets mostly comprise of neutral, sad, happy, and angry emotions.

In the current study, we are using the EMO-DB \cite{burkhardt2005database} dataset, which includes anxiety/fear, sadness, disgust, happiness, anger, neutral, and boredom emotions for our analysis. Additionally, we use the same emotion categories in our analysis of the TESS \cite{dupuis2010toronto} dataset, with the exception of boredom being replaced by surprise. We also expand our experimental investigation with two more datasets, RAVDESS \cite{livingstone2018ryerson} and SAVEE \cite{jackson2014surrey}. These datasets help us evaluate the performance of our approach accurately as they are widely used in the SER field.

\textbf{EMO-DB}:
The Berlin Emotional Speech Database is a collection of 535 German language recordings made by 10 professional actors, 5 male, and 5 female, simulating 7 emotional states: anger, boredom, disgust, anxiety/fear, happiness, sadness, and neutral. The recordings comprise 10 short texts, read in a neutral tone and then in different emotional states. The texts were chosen to be neutral in terms of emotional content to avoid potential emotional carryover effects. Each text is approximately 5 seconds long. The recordings were made in a soundproof studio using a high-quality microphone and recording equipment. The database has been widely used in emotion recognition research and is freely available for non-commercial use.

\textbf{TESS}:
The Toronto Emotional Speech Set is a dataset that comprises 2800 audio recordings. These recordings were made by two actresses, aged 26 and 64, respectively. The actresses were asked to express 200 target phrases in the context of the phrase "Say the word \_\_\_." The dataset includes seven different emotional states, namely, anger, disgust, pleasant surprise, fear, sadness, happiness, and neutral. Each of the seven emotional categories is represented by 400 recordings.

\textbf{RAVDESS}: The Ryerson Audio-Visual Database of Emotional Speech and Song dataset is notable for its wide range of emotions, including happiness, sadness, anger, fear, surprise, and disgust, expressed in speech. It consists of 24 professional actors (12 male, 12 female) who perform each emotion, resulting in a total of 7356 audio-visual recordings. This diversity makes RAVDESS an ideal resource for analyzing the nuances of emotional expression in speech.

\textbf{SAVEE}: The Surrey Audio-Visual Expressed Emotion dataset is distinguished by its focus on male voices, with recordings from 4 male actors enacting different emotions such as anger, disgust, fear, happiness, sadness, surprise, and neutral. It contains 480 utterances, which are valuable for gender-specific emotion recognition studies. The dataset's concentration on a smaller set of actors provides a more controlled environment for analyzing the subtleties of male emotional expression in speech.

\subsubsection{Experimental setup}

\label{exp-setp}

To ensure accurate processing and analysis of audio signals in our SER system, we first establish a sampling rate of 16 KHz in a mono-channel format, aligning with the requirements of most SER algorithms. This step ensures consistency and compatibility with subsequent processing stages. Next, we extract an initial feature set comprising $n=90$ features that are deemed relevant to SER. These features are then normalized and serve as the starting point for our subsequent feature boosting module.

To partition our dataset effectively, we employ stratified random sampling \cite{lyons2018comparison}, maintaining class distribution across three distinct groups: training, validation, and testing. This division ensures representative subsets for training and evaluating our models, with 80\% allocated for training, 10\% for validation, and the remaining 10\% reserved as unseen data for testing.

Moving forward, our feature boosting module becomes instrumental in enhancing the feature representation for SER. It iteratively leverages insights from the explainability module and employs a \textit{VRC-PCA} driven technique to construct a new dataset. This process involves identifying optimal feature combinations that best differentiate emotion categories and capture the most relevant information, reducing dimensionality while preserving crucial discriminative features. The explainability insights guide this selection process, ensuring that the chosen features align with interpretability and contribute significantly to the model's decision-making process. The iterative nature of this module allows us to continuously refine the feature set until convergence is achieved. As a result, we create an improved dataset that reflects the boosted features, offering an enhanced representation of the discriminative patterns underlying emotions. This boosted dataset serves as the foundation for subsequent modules of our method, facilitating improved performance and interpretability in the classification task.

We proceed by training $M=14$ different machine learning models, employing 10-fold cross-validation to assess their performance without overfitting. This technique allows us to evaluate the models' effectiveness and identify the optimal one. Our selection of machine learning models includes Random Forest (RF), Extra Trees classifier (ET), Light Gradient Boosting Machine (LGBM), Linear Discriminant Analysis (LDA), Decision Tree (DT), Quadratic Discriminant Analysis (QDA), Gradient Boosting Classifier (GBC), Logistic Regression (LR), Support Vector Machine (SVM), Naive Bayes (NB), Ridge (Ridge), K-Nearest Neighbors (KNN), Dummy classifier (Dummy), and Adaptive Boosting (ADA). Once the best-performing models have been identified, we employ the grid search technique to fine-tune their hyperparameters. By exhaustively exploring the parameter space, we seek the optimal combination of hyperparameters for each model. This step enhances the robustness and overall performance of the models.

Finally, we evaluate the performance of the final models using the dedicated testing set, ensuring unbiased assessment on unseen data. During the training phase, every segment is employed to predict a single emotion. To ensure the accuracy of our results, we repeat each test 10 times with different random seeds before reporting the average outcome. Additionally, our explainability module incorporates the SHAP approach to evaluate feature importance in the predictions made by the optimal model. This analysis enables us to gain insights into the decision-making process of the model and identify the most influential features for emotion determination. The importance of these features is iteratively communicated back to the feature boosting module, allowing for the refinement of the feature set in subsequent iterations. This iterative process ensures that the most relevant features are retained, while less important ones are discarded, leading to an increasingly optimized feature representation for SER.

\subsubsection{Significance of model explainability}
\label{sig_xai}
\begin{figure}[t]
\begin{minipage}[b]{.47\textwidth}
\includegraphics[width=\textwidth]{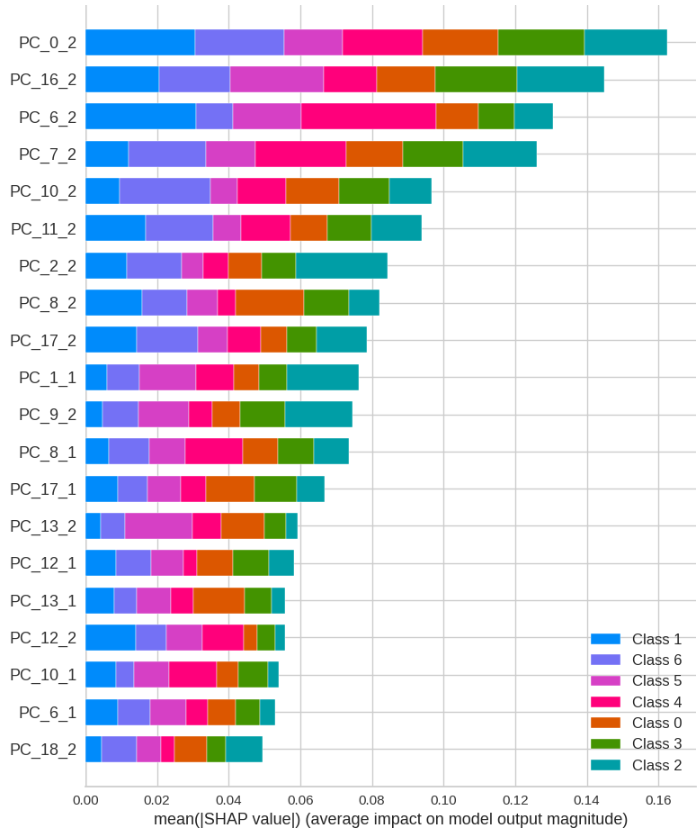}
\caption{Boosted features importance to the model's prediction}\label{fig2}
\end{minipage}\qquad 
\begin{minipage}[b]{.47\textwidth}
\includegraphics[width=\textwidth]{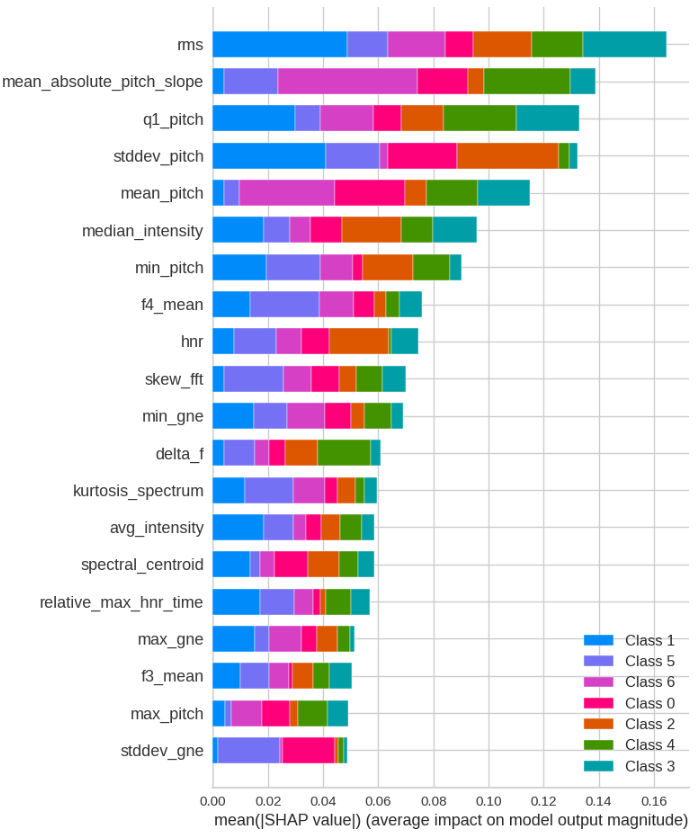}
\caption{Initial features importance to the model's prediction}\label{fig3}
\end{minipage}\\
\centering
\scriptsize{\textit{Class-wise feature importance - Extra Tree classifier (EMO-DB): The x-axis of Fig. \ref{fig2} shows the mean of the absolute values of Shapley values, which indicates the average impact on the model's output magnitude. A higher value suggests a more important feature. The y-axis represents principal components obtained by applying PCA to the optimal combinations of selected features that meet the significance measure (\(\sigma_i\)) threshold. In Fig. \ref{fig3}, the y-axis represents the initial feature importance to the model's decision.}}
\end{figure}

In our study focusing on SER, we place a significant emphasis on integrating XAI techniques into our system. The core idea behind this integration is not only to enhance the performance of our models but also to make them transparent and interpretable. In identifying emotions from speech, it's imperative to understand the 'why' behind the model's predictions. This understanding is key to ensuring the decisions made by our system are both reliable and trustworthy.

To achieve this level of clarity and insight, we use two main strategies: feature importance evaluation and the application of SHAP. Feature importance evaluation is a process where we identify which features in the model are most influential. This could be features like pitch, tone, or speed of speech. Understanding which of these features plays a major role helps us focus on what really matters when it comes to emotion recognition.

The SHAP approach complements this by providing a deeper dive into how each of these important features contributes to the final outcome of the model. SHAP acts as a tool that helps us break down the model's decision process, showing us the contribution of each feature to the prediction. This detailed breakdown is crucial, as it offers a comprehensive understanding of the model's inner workings.

A key aspect of our approach is the integration of a feature boosting module. This module is adept at selecting the most effective feature combinations for emotion recognition. It works hand-in-hand with the explainability module and the classification module, forming a trio that ensures the features we select are not just relevant but also contribute to making the model more understandable. This is crucial because it means our model isn't just a black box; it's a system whose decisions can be traced and understood.

\begin{figure}[t]
\centerline{\includegraphics[width=.6\textwidth]{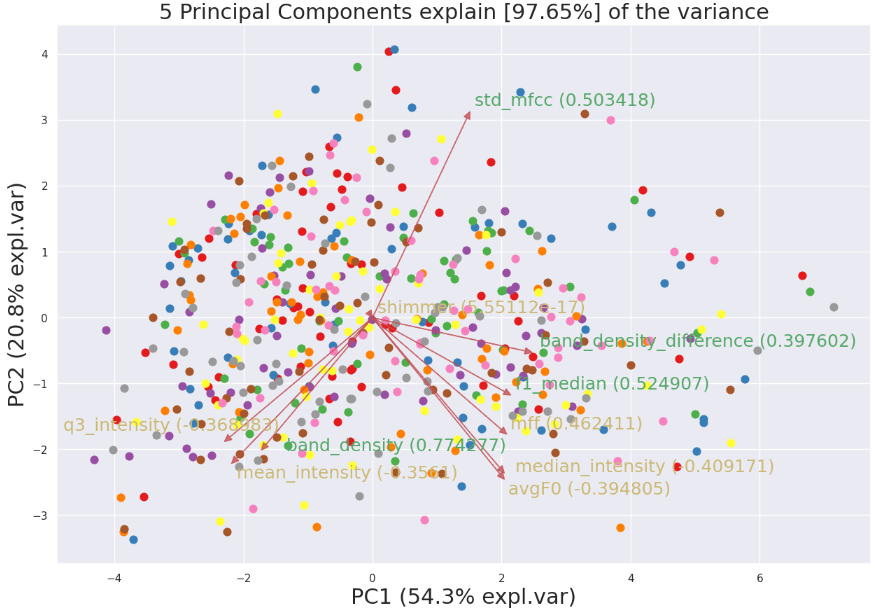}}
\caption{Biplot of the $2^{nd}$ Feature combination (EMO-DB)}
\label{fig4}
\end{figure}

Our approach's effectiveness is demonstrated through visualizations in Figs. \ref{fig2} and \ref{fig3}, where we show how each feature contributes to predicting different emotions. For instance, in Fig. \ref{fig2}, we label principal components in a way that allows us to easily track which features are most influential in emotion classification. The principal components are labeled as $PC_{\{combination\_index\}\{PC\_index\}}$, where $PC_{ij}$ refers to the $j^{th}$ principal component of the $i^{th}$ feature combination. This labeling is instrumental in identifying key components and understanding their role in the model's predictions.

Analyzing Fig. \ref{fig2}, we can identify the principal components that have the highest impact on the model's output. Additionally, Fig. \ref{fig3} helps us determine the contribution of each initial feature to the final decision. Fig. \ref{fig4} takes this a step further by providing a biplot (e.g. $2^{nd}$ feature combination). This visualization helps us see not only the principal components that are most impactful but also how the initial features contribute to these components. This level of detail is important as it guides us in iteratively eliminating less relevant features and refining our model. By continuously identifying and focusing on the most relevant features, we enhance the accuracy of our SER system.

In summary, the integration of feature boosting and a thorough analysis of feature contributions are pivotal in our SER framework. It enables us to identify the most relevant features for emotion recognition, thereby enhancing the accuracy and reliability of our classification decisions. More importantly, it offers a window into our model's thought process, ensuring that each decision made by our system is transparent and understandable.

\subsubsection{Evaluation metrics}
\label{eval-mtrc}
For performance evaluation, we use accuracy, recall, precision, and F1-Score, which attempts to establish a compromise between precision and recall metrics {\cite{tharwat2020classification}. By using these metrics, we can measure the effectiveness of our approach and compare its performance with other state-of-the-art methods.

\subsection{Results}

\begin{table}[t]
\centering
\caption{Features comprised in EMO-DB optimal combination}

\begin{tabular}{ll}
\hline
\textit{\textbf{Feature abbreviation}} & \textbf{Feature full name and significance}\\ \hline
\textbf{q1\_fft} & First quartile of Fast Fourier Transform magnitudes \\ \hline
\textbf{mff} & Mel Frequency Flux \\ \hline
\textbf{median\_fft} & Median magnitude of the Fast Fourier Transform \\ \hline
\textbf{q1\_pitch} & First quartile of pitch values \\ \hline
\textbf{q3\_pitch} & Third quartile of pitch values \\ \hline
\textbf{mean\_fft} & Mean magnitude of the Fast Fourier Transform \\ \hline
\textbf{minv\_mfcc} & Minimum value of Mel Frequency Cepstral Coefficients \\ \hline
\textbf{shimmer} & Amplitude variation in consecutive voice cycles \\ \hline
\textbf{mean\_pitch} & Average pitch value of the speech signal \\ \hline
\textbf{skew\_mfcc} & Skewness of Mel Frequency Cepstral Coefficients \\ \hline
\end{tabular}

\label{table1}
\end{table}

\subsubsection{Feature boosting}

In our study, we apply a threshold on $\sigma_i$, by fixing the value of $\epsilon$, this allows us to select only the feature combinations that best separate emotion classes with optimal density resulting in 23 optimal combinations for TESS dataset and 25 for EMO-DB dataset. When applying PCA, on TESS dataset, one of these combinations captures 84.56\% of the total explained variance in the first two principal components and 98.22\% in the first four. On EMO-DB dataset, the explained variance by the first two principal components of the optimal feature set to represent the information in the data is 76.8\% and 99.36\% by the first five principal components, making it the best feature set to represent the information in the data (e.g. see Table \ref{table1}). This reflects the amount of information explained by the principal components of the retained feature combinations.

\begin{figure}[t]
\centerline{\includegraphics[width=.6\textwidth]{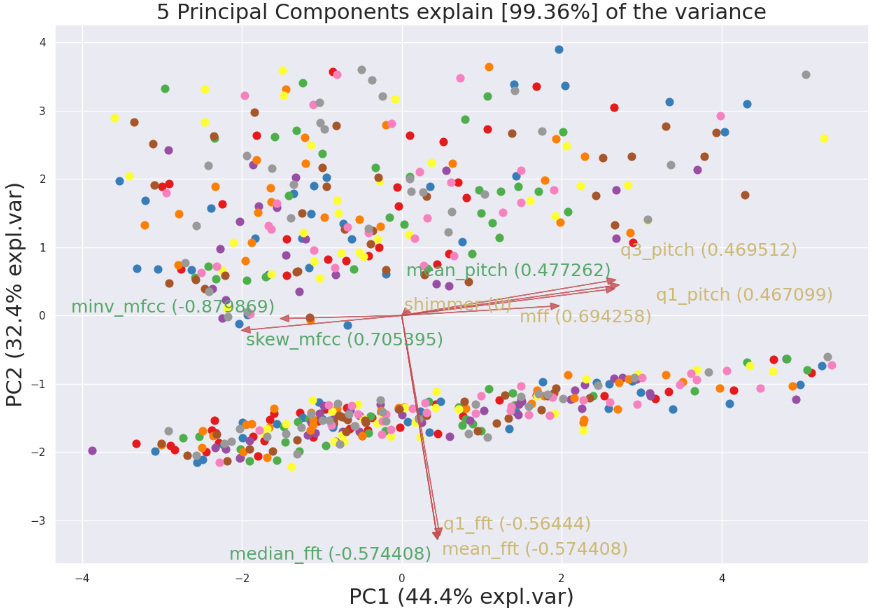}}
\caption{Biplot of EMO-DB optimal feature combination}
\scriptsize{\textit{The magnitude of the arrows indicates the significance of each feature within the dataset, while the angle between the arrows signifies the correlation between the features.}}
\label{fig5}
\end{figure}

\begin{figure}[t]
\centerline{\includegraphics[width=.6\textwidth]{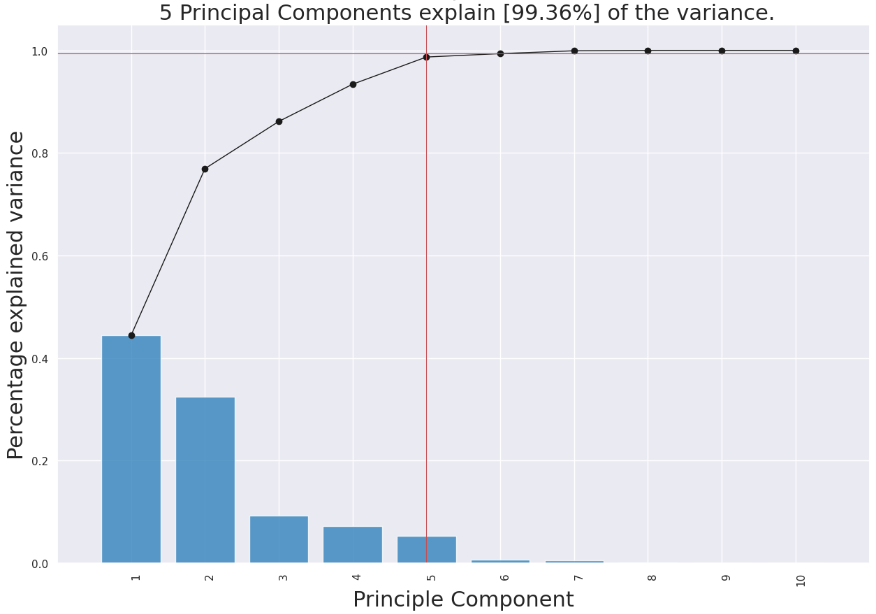}}
\caption{Cumulative explained variance of EMO-DB optimal feature combination}
\label{fig6}
\end{figure}

To visualize the relationship between the optimal features and principal components, we use a biplot as shown in Fig. \ref{fig5}, which displays the data points on a 2D scatter plot based on the values of the first two principal components of the optimal feature combination data. The biplot also shows the directions and lengths of arrows representing the optimal features in the transformed space. The direction of the arrow indicates the sign of the contribution, while the length indicates the magnitude of the contribution. This allows us to understand how the optimal features are related to the principal components and how they contribute to the overall variance in the data.

Furthermore, we add the cumulative explained variance in Fig. \ref{fig6} to the biplot, which shows the percentage of the total variance in the optimal feature combination data explained by each principal component. This helps us determine the optimal number of principal components to retain when performing PCA on the selected feature combinations of each dataset for a convenient information representation.

\subsubsection{Importance of feature boosting and model explainability}\hfill \break
\textbf{On TESS dataset:}

\begin{table}[t]
\centering
\caption{Compared models on all initially computed features in (\%) on TESS dataset: Best results are in bold font}

\begin{tabular}{lllll}
\hline
\textit{\textbf{Model}}   & \textbf{Accuracy} & \textbf{Recall}  & \textbf{Precision}  & \textbf{F1-score}\\ \hline
\textbf{ET}      & \textbf{95.8}  & \textbf{95.8} & \textbf{95.8}  & \textbf{95.8}  \\ \hline
\textbf{LGBM}     &  95 &	94.9 &	95 &	94.7  \\ \hline
\textbf{RF}       & 94.6 &	94.6 &	94.7 &	94.5  \\ \hline
\textbf{KNN}      & 94.4 &	94.4 &	97.3 &	97.4  \\ \hline
\textbf{GBC}      & 94.3 &	94.3 &	94.3 &	94.3  \\ \hline
\textbf{LR}       & 94 &	94 &	94.1 &	93.9  \\ \hline
\textbf{LDA}      & 92.9 &	93 &	93.3 & 92.9  \\ \hline
\textbf{DT}     & 92.9 &	92.1 &	92.3 &	92.1  \\ \hline
\textbf{SVM}    & 91.6 &	91.7 &	92.2 &	91.6 \\ \hline
\textbf{Ridge}   &  88.4 &	88.5 &	89.3 &	87.9 \\ \hline
\textbf{NB}  & 85.4 & 85.4 &	86.1 &	85.2  \\ \hline
\textbf{QDA}   & 83.2 &	83.3 &	85.7 &	82  \\ \hline
\textbf{ADA}   & 51.7 &	51.9 &	50.4 &	45.4 \\ \hline
\textbf{Dummy}  & 14.6 &	14.3 &	2.1 &	3.7  \\ \hline
\end{tabular}

\label{table2}
\end{table}

\begin{figure}[t]
\centerline{\includegraphics[width=.7\textwidth]{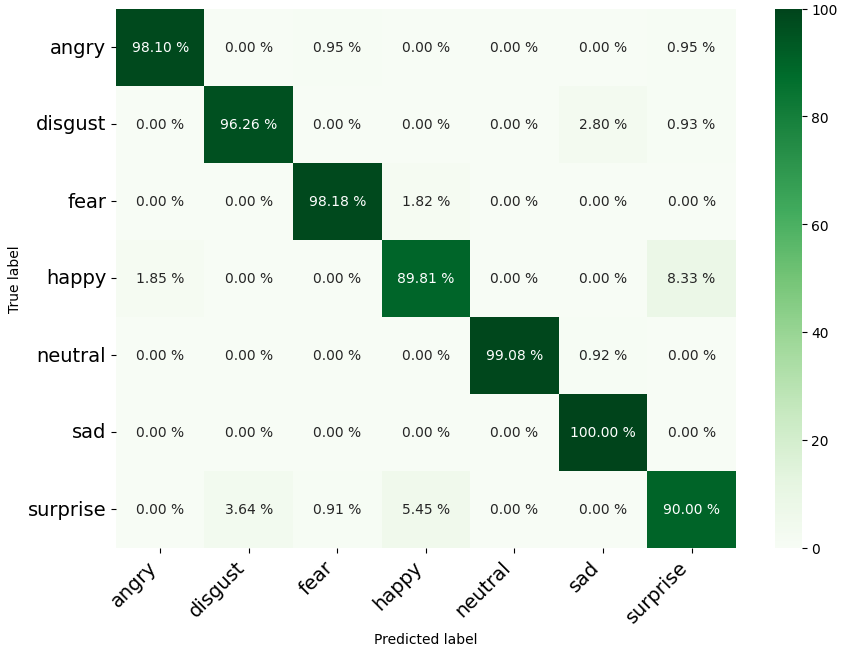}}
\caption{Extra Tree classifier confusion matrix without feature boosting and model explainability feedback loop on TESS dataset}
\label{fig7}
\end{figure}

\begin{table}[t]
\centering
\caption{Compared models on the constructed dataset using feature boosting and model explainability feedback loop in (\%) on TESS dataset: Best results are in bold font}

\begin{tabular}{lllll}
\hline
\textit{\textbf{Model}}   & \textbf{Accuracy} & \textbf{Recall}  & \textbf{Precision}  & \textbf{F1-score}\\ \hline

\textbf{ET}		& \textbf{99.4}	& \textbf{99.9}	& \textbf{99.4}	& \textbf{99.4} \\ \hline
\textbf{LGBM}		& 99  &  99 &	99 &  99\\ \hline
\textbf{RF}	& 	98.9 &  98.9 &	98.9 &  98.9 \\ \hline
\textbf{QDA}	& 98.6 &  98.7 &	98.7 &  98.6 \\ \hline
\textbf{GBC}   & 98.6 &  98.6 &	98.7 &  98.7 \\ \hline
\textbf{LDA}	& 98.5 &  98.5 &	98.6 &  98.5 \\ \hline
\textbf{DT}		& 93.3 &  93.3 &	93.6 &  93.2 \\ \hline
\textbf{Ridge}	& 50.1 &  50 &	59.8 &  50.9 \\ \hline
\textbf{ADA}	& 39.4 &  39.2 &	39.2 &  30.9 \\ \hline
\textbf{NB}		& 24  &  23.6 &	13.4 &  16.2 \\ \hline
\textbf{KNN}	& 22.9  &  22.7 &	21.8 &  21.6 \\ \hline
\textbf{Dummy}	& 14.6 &  14.3 &	2.1 &  3.7 \\ \hline
\textbf{LR}		& 14.5 &  14.3 &	2.1 &  3.7 \\ \hline
\textbf{SVM}	& 14.5 &  14.3 &	2.1 &  3.7 \\ \hline
\end{tabular}

\label{table3}
\end{table}

\begin{figure}[t]
\centerline{\includegraphics[width=.7\textwidth]{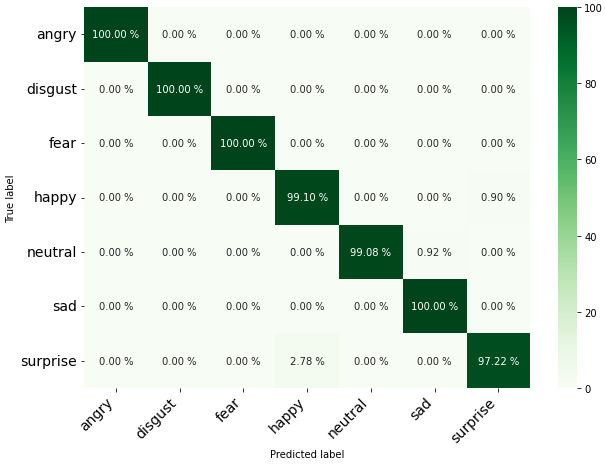}}
\caption{Extra Tree classifier confusion matrix with feature boosting and model explainability feedback loop on TESS dataset}
\label{fig8}
\end{figure}

Table \ref{table2} presents the performance of the selected machine learning models on the initially computed features for the SER task using TESS dataset. The ET classifier achieves the highest accuracy, recall, precision, and F1-score of 95.8\%. The LGBM also performs well with 95\% accuracy. RF and KNN have a relatively high accuracy of 94.6\% and 94.4\%, respectively. These results suggest that the initially computed features contain valuable information for the SER task. The confusion matrix in Fig. \ref{fig7} shows that the ET classifier performs well overall, correctly predicting the diagonal elements of each class. However, there are some misclassifications, indicating that some classes share acoustic similarities. 

Table \ref{table3} compares the performance of the same models using boosted features with explainability module feedback loop. The ET classifier achieves the highest accuracy and F1-score of 99.4\%. The LGBM also performs well with an accuracy and F1-score of 99\%. RF has a relatively high accuracy of 98.9\%. Some models perform poorly and lose effectiveness when feature boosting and model explainability are used, indicating that they may be less suitable for SER applications than the well-performing models. In conclusion, the ET classifier and LGBM are the best-performing models for TESS dataset, achieving high accuracy and F1-score. The confusion matrix shown in Fig. \ref{fig8} displays the performance of the ET classifier, which performs well overall with a high number of correct predictions on the diagonal elements for each emotion. For instance, all actual "angry" classes are correctly predicted as "angry" and all actual "disgust" classes are correctly predicted as "disgust". However, there are still some misclassifications, such as 0.9\% of "happy" being predicted as "surprise", 0.92\% of "neutral" being predicted as "sad", and 2.78\% of "surprise" being predicted as "happy". This suggests that "happy" and "neutral" share some characteristics with "surprise" and "sad", respectively.

To optimize model performance, hyperparameter tuning is performed by finding the best values for the hyperparameters that control the model's complexity and generalization performance, as is the case with our compared models. We use the random grid search technique to achieve this, training the model using a range of hyperparameter values and evaluating the performance of each model using cross-validation sets. The best set of hyperparameters is then selected based on the model's performance on the validation set.

\hfill \break
\textbf{On EMO-DB dataset:}

\begin{table}[t]
\centering
\caption{Compared models on all initially computed features in (\%) on EMO-DB dataset: Best results are in bold font}

\begin{tabular}{lllll}
\hline
\textit{\textbf{Model}}   & \textbf{Accuracy} & \textbf{Recall}  & \textbf{Precision}  & \textbf{F1-score}\\ \hline
\textbf{ET}     &  \textbf{75.3} &	\textbf{71.5} &	\textbf{77.4} &	\textbf{73.9} \\ \hline
\textbf{LGBM}       & 64.6 &	61.2 &	66.9 &	63.5  \\ \hline
\textbf{LR}      & 63.1 &	60 &	62.6 &	61.4  \\ \hline
\textbf{RF}      & 62.8 &	59.4 &	64.4 &	60.7  \\ \hline
\textbf{GBC}       & 62.2 &	58.7 &	62.9 &	61  \\ \hline
\textbf{LDA}      & 60.7 &	58.5 &	63.4 & 59.8  \\ \hline
\textbf{Ridge}     & 60.4 &	55.6 &	59.6 &	56.8  \\ \hline
\textbf{KNN}    & 56.3 & 50.8 &	55.7 &	52.7 \\ \hline
\textbf{SVM}   &  54.7 &	51.9 &	57.3 &	53.1 \\ \hline
\textbf{QDA}  & 51.1 & 43.1 &	50 &	47.1  \\ \hline
\textbf{DT}   & 50.3 &	47.9 &	50.9 &	48.9  \\ \hline
\textbf{NB}   & 49.4 &	46.1 &	50.7 &	47.6 \\ \hline
\textbf{ADA}  & 37.5 &	31.9 &	34.4 &	31.5  \\ \hline
\textbf{Dummy}  & 23.5 &	14.3 &	5.5 & 9  \\ \hline
\end{tabular}

\label{table4}
\end{table}

Table \ref{table4} compares the performance of the selected machine learning models on the initially computed features for the SER task using the EMO-DB dataset. The ET classifier outperforms other models with the highest accuracy, recall, precision, and F1-score of 75.3\%. The LGBM model achieves the second-highest accuracy of 64.6\%, followed by LR with an accuracy of 63.1\%. The RF and GBC models have a slightly lower accuracy of 62.8\% and 62.2\%, respectively. The LDA model has an accuracy of 60.7\%. The performance of the remaining models, including Ridge, KNN, SVM, QDA, DT, ADA, and Dummy, are lower with accuracy scores ranging from 23.5\% to 56.3\%. These results indicate that the designed features have a moderate impact on the SER task using the EMO-DB dataset, and the ET model achieves the best overall performance among the compared models.

\begin{figure}[t]
\centerline{\includegraphics[width=.7\textwidth]{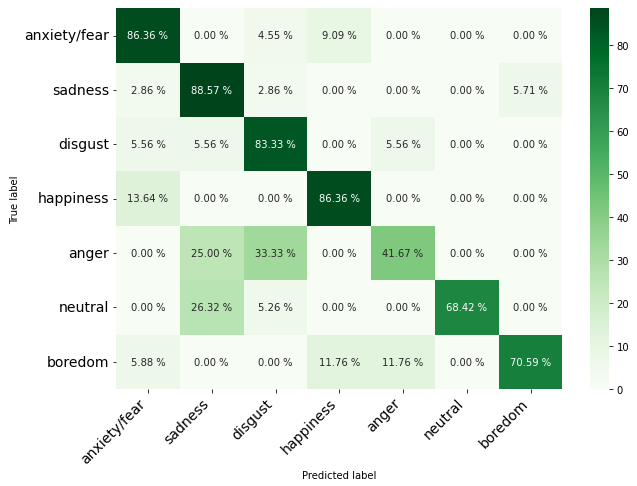}}
\caption{Extra Tree classifier confusion matrix without feature boosting and model explainability feedback loop on EMO-DB dataset}
\label{fig9}
\end{figure}

\begin{table}[t]
\centering
\caption{Compared models on the constructed dataset using feature boosting and model explainability feedback loop in (\%) on EMO-DB dataset: Best results are in bold font}

\begin{tabular}{lllll}
\hline
\textit{\textbf{Model}}   & \textbf{Accuracy} & \textbf{Recall}  & \textbf{Precision}  & \textbf{F1-score}\\ \hline
\textbf{ET}      & \textbf{88.3}  & \textbf{86.8} & \textbf{88.6}  & \textbf{87.4}  \\ \hline
\textbf{LGBM}     &  73.8 &	69.4 &	74 &	70.5  \\ \hline
\textbf{LDA}       & 73.2 &	71.2 &	75.9 &	72.6  \\ \hline
\textbf{RF}      & 70 &	65.1 &	70.6 &	67.3  \\ \hline
\textbf{GBC}      & 69 &	65.4 &	72.3 &	68.4  \\ \hline
\textbf{Ridge}       & 62 &	56.6 &	61.1 &	58.9  \\ \hline
\textbf{DT}      & 50.6 & 48 &	52 & 49.3  \\ \hline
\textbf{ADA}     & 37.5 &	31 &	27.2 &	28.2  \\ \hline
\textbf{NB}    & 27.7 &	20 &	12.6 &	15.8 \\ \hline
\textbf{KNN}   &  24.1 & 21.6 &	24.5 &	22.9 \\ \hline
\textbf{Dummy}  & 23.5 & 14.3 &	5.5 &	9  \\ \hline
\textbf{QDA}   & 23.2 &	17.1 &	14.2 &	15.2  \\ \hline
\textbf{SVM}   & 13.1 &	14.3 &	1.9 &	3.2 \\ \hline
\textbf{LR}  & 11.6 &	14.3 &	1.4 &	2.4  \\ \hline
\end{tabular}

\label{table5}
\end{table}

The confusion matrix in Fig. \ref{fig9} represents the performance of the ET classifier on EMO-DB. Looking at the diagonal values, we can see that the model performs well on some emotions such as "sadness" achieving 88.57\% accuracy, and "anxiety/fear" and "happiness" achieving 86.36\% accuracy for both. However, the model performs less with other emotions such as "disgust", "anger", "neutral" and "boredom" achieving 83.33\%, 41.67\%, 68.42\%, and 70.59\% accuracy respectively. There are also some off-diagonal values, indicating misclassifications. For example, the model seems to frequently misclassify "boredom" as "anger" or "happiness" (11.76\% of the instances each), which could indicate some similarity between these emotions. Additionally, the model has difficulty distinguishing between "anger" and both "sadness" and "disgust", with it being misclassified as "sadness" in 25\% of the instances and as "disgust" in 33\% of the instances. While the model performs well on some emotions, it slightly struggles with others and there is room for improvement.

Table \ref{table5} shows the performance of the selected machine learning models on the constructed dataset using feature boosting and model explainability feedback loop, based on the EMO-DB dataset. The best-performing model is the ET model, which achieves an accuracy of 88.3\%, recall of 86.8\%, precision of 88.6\%, and F1 score of 87.4\%. The second-best model is the LGBM model, which achieves an accuracy of 73.8\%, recall of 69.4\%, precision of 74\%, and F1 score of 70.5\%. The other models, such as LDA, RF, GBC, Ridge, DT, ADA, NB, KNN, Dummy, QDA, SVM, and LR, achieve lower performance than the top two models, with accuracy scores ranging from 62\% to as low as 11.6\%. The worst-performing models based on all the metrics are LR and SVM models, with accuracy scores of 11.6\% and 13.1\%, respectively. Overall, the results suggest that the ET and LGBM models are the optimal options for the EMO-DB dataset.

\begin{figure}[t]
\centerline{\includegraphics[width=.7\textwidth]{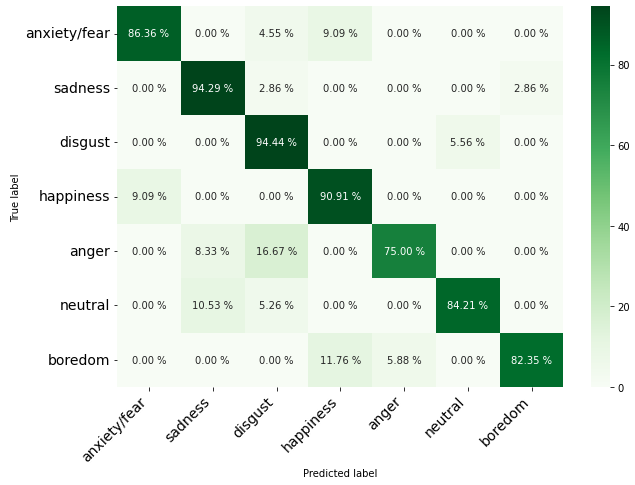}}
\caption{Extra Tree classifier confusion matrix with feature boosting and model explainability feedback loop on EMO-DB dataset}
\label{fig10}
\end{figure}

This confusion matrix in Fig. \ref{fig10} shows the performance of ET model in classifying different emotions in the EMO-DB dataset. Compared to the previous confusion matrix in Fig. \ref{fig9}, this one shows an improvement in overall accuracy and precision, as well as some changes in the patterns of misclassifications. For example, the model is improved in classifying "sadness" correctly, with 94.29\% accuracy, but there is still some confusion between "neutral" and "sadness" emotions, with 10.53\% of "neutral" emotions being misclassified as "sadness". Similarly, the model is improved in classifying "disgust" correctly, with 94.44\% accuracy, but there is still some confusion between "neutral" and "disgust" emotions, with 5.56\% of "neutral" emotions being misclassified as "disgust". The model is also improved in classifying "anger" correctly, with 75\% accuracy, but there is still some confusion between "anger" and "disgust", with 16.67\% of "anger" being misclassified as "disgust". We assume that these missclassifications are due to a remarkable acoustic similarity between the mutually missclassified emotions. However, the model shows a considerable improvement in correctly classifying most of the emotions when using our approach. 

In summary, our experiments demonstrate the significant improvements achieved by integrating feature boosting and model explainability techniques into our SER system. Through dimensionality reduction, noise removal, and redundancy elimination, the models benefit from a more informative representation of the data, leading to enhanced performance and robustness. The explainability module plays a crucial role in identifying key feature combinations that drive the classification process. This iterative feature selection process further enhances the representation of the speech signal and improves the generalization performance of the models. By incorporating feature boosting and model explainability techniques, we achieve a comprehensive and refined approach that maximizes the utilization of relevant information while minimizing noise and redundancy, resulting in a more effective SER system.

Our results highlight the effectiveness of the ET classifier and LGBM for the SER task, as they achieved high accuracy and F1-score. While the initially computed features are valuable for representing speech signals and capturing emotional content, the boosted features outperform them in distinguishing between emotions. The observed misclassifications in the confusion matrices indicate that some emotions may share similar acoustic features, posing challenges in differentiation. However, the feedback mechanism between feature boosting and model explainability proves effective in enhancing model performance and robustness.

Additionally, hyperparameter tuning techniques, such as random grid search, optimize model performance by finding the best values for controlling model complexity and generalization. This further enhances the robustness and performance of the SER system. Overall, our study demonstrates the value of integrating feature boosting and model explainability in developing an advanced and reliable SER system capable of accurately recognizing emotions in speech signals.

\subsubsection{Comparison with SOTA methods}\hfill \break
\label{comp_sota}
\textbf{On TESS dataset:}

\begin{table}[t]
    \centering
    \caption{Compared methods performance on TESS dataset: best results are in bold font}
    \begin{tabular}{lll}
        \hline
        \multicolumn{3}{c}{\textbf{Compared methods}}                                           \\ \hline
        \textit{\textbf{TESS dataset}}              & \textbf{Test Accuracy (\%)} & \textbf{F1-score (\%)} \\ \hline
        \textbf{Aggarwal et al. \cite{aggarwal2022two}} & 97.6  &  97 \\ \hline
        \textbf{Praseetha et al. \cite{praseetha2018deep}} &    95.8 &  NA    \\ \hline
        \textbf{Choudhary et al. \cite{choudhary2022speech}}     &   97.1 &  96 \\ \hline
        \textbf{Iqbal et al. \cite{Iqbal2020MFCCAM}} &   97 &  NA  \\ \hline
        \textbf{Kapoor et al. \cite{kapoor2022fusing}}  &   97.5  &  97.4  \\ \hline
        \textbf{Krishnan et al. \cite{krishnan2021emotion}}  &   93.3  &  NA   \\ \hline
        \textbf{Stawicki et al. \cite{stawicki2024ensembles}} &    96.5 &  96.7  \\ \hline
        \textbf{Dupuis et al. (HLP) \cite{dupuis2011recognition}} &    82 &  NA  \\ \hline
        \textbf{Our method} &  \textbf{99.4}  &  \textbf{99.4} \\ \hline
        \end{tabular}
      \label{table6}
\end{table}

To further validate our findings, our proposed method's performance on the TESS dataset is compared to other state-of-the-art methods, as shown in Table \ref{table6}. We use two main evaluation approaches for our method. 

First, we compare our method against human-level performance (HLP) on the TESS dataset, as evaluated in \cite{dupuis2011recognition}, where authors used 56 human annotators to recognize emotions. Second, we compare our method against machine learning-based SER methods. As previously discussed, SERs typically involve two main stages: feature extraction and classification. Many of the compared methods in the literature used MFCC for feature extraction, such as \cite{Iqbal2020MFCCAM}, \cite{praseetha2018deep}, and \cite{choudhary2022speech}, while some others used spectrograms combined with Empirical Mode Decomposition (EMD) \cite{krishnan2021emotion} or PCA \cite{aggarwal2022two}. For classification, some methods employed traditional machine learning techniques such as SVM \cite{Iqbal2020MFCCAM}, Latent Dirichlet Allocation \cite{krishnan2021emotion} or decision bireducts \cite{stawicki2024ensembles} as an extension of decision reducts in rough set theory, offering a rule-based classification.  While others used deep neural networks \cite{aggarwal2022two}, \cite{choudhary2022speech}, \cite{praseetha2018deep}, and \cite{kapoor2022fusing}.

Our proposed method achieves an accuracy of 99.4\% and an F1-score of 99.4\%, which are the highest scores among all the compared methods. For HLP, an accuracy of 82\% is achieved on the TESS dataset. While for the machine learning based method, an accuracy of 97.6\% and an F1-score of 97\% are achieved by \cite{aggarwal2022two}, \cite{kapoor2022fusing} achieved an accuracy of 97.5\% and an F1-score of 97.4\%, and \cite{choudhary2022speech} achieved an accuracy of 97.1\% and an F1-score of 96\%. \cite{praseetha2018deep}, \cite{Iqbal2020MFCCAM}, and \cite{krishnan2021emotion} achieved accuracies of 95.8\%, 97\%, and 93.3\%, respectively.

In summary, our proposed method outperforms all the compared state-of-the-art machine learning-based SER methods and it also achieves a performance that exceeds human-level performance (HLP) on the TESS dataset.

\noindent\textbf{On EMO-DB dataset:}

\begin{table}[t]
    \centering
    \caption{Compared methods performance on EMO-DB dataset: best results are in bold font}
    \begin{tabular}{lll}
        \hline
        \multicolumn{3}{c}{\textbf{Compared methods}}                                           \\ \hline
        \textit{\textbf{EMO-DB dataset}}              & \textbf{Test Accuracy (\%)} & \textbf{F1-score (\%)} \\ \hline
        \textbf{Pham et al. \cite{pham2021emotion}} &    76.4 &  NA    \\ \hline
        \textbf{Ancilin et al. \cite{ancilin2021improved}} & 81.5 &  NA \\ \hline
        \textbf{Singh et al. \cite{singh2020efficient}} & 86.36 & NA \\ \hline
        \textbf{Seo et al. \cite{seo2020fusing}} & 86.92 & 86.7 \\ \hline
        \textbf{Stawicki et al. \cite{stawicki2024ensembles}} &    85.8 &  85.6  \\ \hline
        \textbf{Mustaqeem et al. \cite{sajjad2020clustering}} & 85.57 & 85 \\ \hline
        \textbf{Our method} &  \textbf{88.3}  &  \textbf{87.4} \\ \hline
        \end{tabular}
      \label{table7}
\end{table}

To compare our proposed method against state-of-the-art approaches, we select recent studies that focus on SER using machine learning techniques. The first study, \cite{pham2021emotion}, focuses on deep learning for SER using CNNs on the EMO-DB dataset. The authors use different spectral features for acoustic signal collections and obtain unweighted average accuracy values of 99.3\% and 76.4\% on the two-class and seven-class EMO-DB datasets, respectively. The second approach presented in \cite{ancilin2021improved} proposes an improved method for SER using the Mel frequency magnitude coefficient as the feature. The authors test the proposed method on several databases, including EMO-DB, and report an accuracy of 81.5\% on the EMO-DB dataset using multiclass SVM as the classifier. Authors in \cite{singh2020efficient} leveraged optimized feature selection and classifier tuning to significantly enhance language-independent SER accuracy achieving an accuracy of 86.36\%. In \cite{sajjad2020clustering}, a framework that fuses key sequence segment selection, CNN-based deep feature extraction, and Bi-LSTM-driven temporal information learning is proposed, achieving an accuracy of 85.57\% and an F1-score of 85\%. The compared methods are presented in Table \ref{table7}, where the best results on the EMO-DB dataset are highlighted in bold font. According to these results, our proposed method outperforms these methods with an accuracy of 88.3\% and an F1-score of 87.4\%.

\subsubsection{Performance and comparative analysis with SOTA methods across additional datasets: RAVDESS, SAVEE}\label{new_data_comp_sot}

The additional datasets utilized in the comparative analysis of SER methods each provide a further unique landscape for evaluating the effectiveness of these technologies, enriched by their diverse emotional content and specific characteristics. Each dataset, with its distinct set of emotional expressions and contexts, forms a comprehensive base for evaluating and advancing SER methods. The variety in the types of emotions, the modes of expression, and the demographic diversity of the actors across these datasets offer a rich, multifaceted perspective crucial for robust SER technology development.

In the comparative analysis of our method with other state-of-the-art methods in the field of SER across different datasets, we observe notable trends and performances. Our method showcases a leading performance on the RAVDESS dataset with accuracy and F1-score both at 87.3\%, outperforming other techniques such as \cite{farooq2020impact} at 81.3\%, and \cite{er2020novel} at 79.41\%. Notably, some methods like \cite{singh2020efficient} yielded a lower accuracy of 64.15\%, as shown in Table \ref{table8}. This variation in performance highlights the efficacy of our approach in handling the complexities of the RAVDESS dataset. Our approach again leads on the SAVEE dataset, as we can see in Table \ref{table9}, demonstrating a high performance with an accuracy of 85.4\% and an F1-score of 85.5\%, outperforming \cite{farooq2020impact} which achieved 82.1\% accuracy and 82.3\% F1-score, and \cite{singh2020efficient} at 77.38\% accuracy. This trend further validates the effectiveness of our method in accurately recognizing emotions from speech, as this consistently high performance across different datasets underlines the robustness and adaptability of our method.

Across all datasets, our method consistently achieved better performance, underscoring its effectiveness in SER tasks. This superior performance across various datasets indicates a robust and adaptable approach, capable of handling different emotional contexts and speech nuances effectively. The comparison with other methods, ranging from traditional machine learning techniques to advanced deep learning models with different feature selection approaches, reveals the importance of carefully boosting the features used in SER.

\begin{table}[t]
    \centering
    \caption{Compared methods performance on RAVDESS dataset: best results are in bold font}
    \begin{tabular}{llll}
        \hline
        \multicolumn{4}{c}{\textbf{RAVDESS Dataset}}                                           \\ \hline
        \textbf{Study} & \textbf{Method} & \textbf{Test Acc. (\%)} & \textbf{F1. (\%)} \\ \hline
        \textbf{Farooq et al. \cite{farooq2020impact}} & DCNN with various classifiers & 81.3 & NA \\ \hline
        \textbf{Er et al. \cite{er2020novel}} & Acoustic and deep features, SVM & 79.41 & NA \\ \hline
        \textbf{Singh et al. \cite{singh2020efficient}} & Feature and classifier optimization & 64.15 & NA \\ \hline
        \textbf{Kanwal et al. \cite{kanwal2021speech}} & Clustering-based genetic algorithm & 82.5 & NA \\ \hline
        \textbf{Radoi et al. \cite{radoi2021end}} & End-to-end neural network TA-AVN & 78.7 & NA \\ \hline
        \textbf{Ezz-Eldin et al. \cite{ezz2021efficient}} & Hybrid CNN and feedforward DNN & 80.6 & 81.1 \\ \hline
        \textbf{Xu et al. \cite{xu2021head}} & Attention-based ACNN model & 76.18 & NA \\ \hline
        \textbf{Ancilin \& Milton \cite{ancilin2021improved}} & Mel frequency magnitude coefficient & 64.31 & NA \\ \hline
        \textbf{Seo et al. \cite{seo2020fusing}} & Visual attention CNN and BOVW & 83.33 & 83.4 \\ \hline
        \textbf{Stawicki et al. \cite{stawicki2024ensembles}} & Decision bireducts based classification  &  85.3 &  85.2  \\ \hline
        \textbf{Aggarwal et al. \cite{aggarwal2022two}} & Two-way feature extraction and VGG-16 & 81.94 & NA \\ \hline
        \textbf{Mustaqeem et al. \cite{sajjad2020clustering}} & Clustering-Based with BiLSTM & 77.02 & 77 \\ \hline
        \textbf{Mocanu et al. \cite{mocanu2021utterance}} & SE-ResNet with spectrogram inputs & 83.35 & NA \\ \hline
        \textbf{Our method} & Iterative feature boosting & \textbf{87.3}  &  \textbf{87.3} \\ \hline
    \end{tabular}
    \label{table8}
\end{table}

\begin{table}[t]
    \centering
    \caption{Compared methods performance on SAVEE dataset: best results are in bold font}
    \begin{tabular}{llll}
        \hline
        \multicolumn{4}{c}{\textbf{SAVEE Dataset}}                                           \\ \hline
        \textbf{Study} & \textbf{Method} & \textbf{Test Acc. (\%)} & \textbf{F1. (\%)} \\ \hline
        \textbf{Farooq et al. \cite{farooq2020impact}} & DCNN with Feature Selection Algorithm & 82.1 & 82.3 \\ \hline
        \textbf{Singh et al. \cite{singh2020efficient}} & Feature and classifier optimization & 77.38 & NA \\ \hline
        \textbf{Kanwal et al. \cite{kanwal2021speech}} & Clustering-based genetic algorithm & 77.7 & NA \\ \hline
        \textbf{Seo et al. \cite{seo2020fusing}} & Visual attention CNN and BOVW & 75 & 75.2 \\ \hline
        \textbf{Stawicki et al. \cite{stawicki2024ensembles}} &  Decision bireducts based classification & 79.4 &  79.5 \\ \hline
        \textbf{Amjad et al. \cite{amjad2022recognizing}} & 1D and 2D DCNNs with LSTM & 77.64 & NA \\ \hline
        \textbf{Ancilin et al. \cite{ancilin2021improved}} & Mel frequency magnitude coefficient &  75.63 & NA \\ \hline
        \textbf{Our method} & Iterative feature boosting & \textbf{85.4}  &  \textbf{85.5} \\ \hline
    \end{tabular}
    \label{table9}
\end{table}

\begin{figure}[t]
\begin{minipage}[b]{.5\textwidth}
\includegraphics[width=\textwidth]{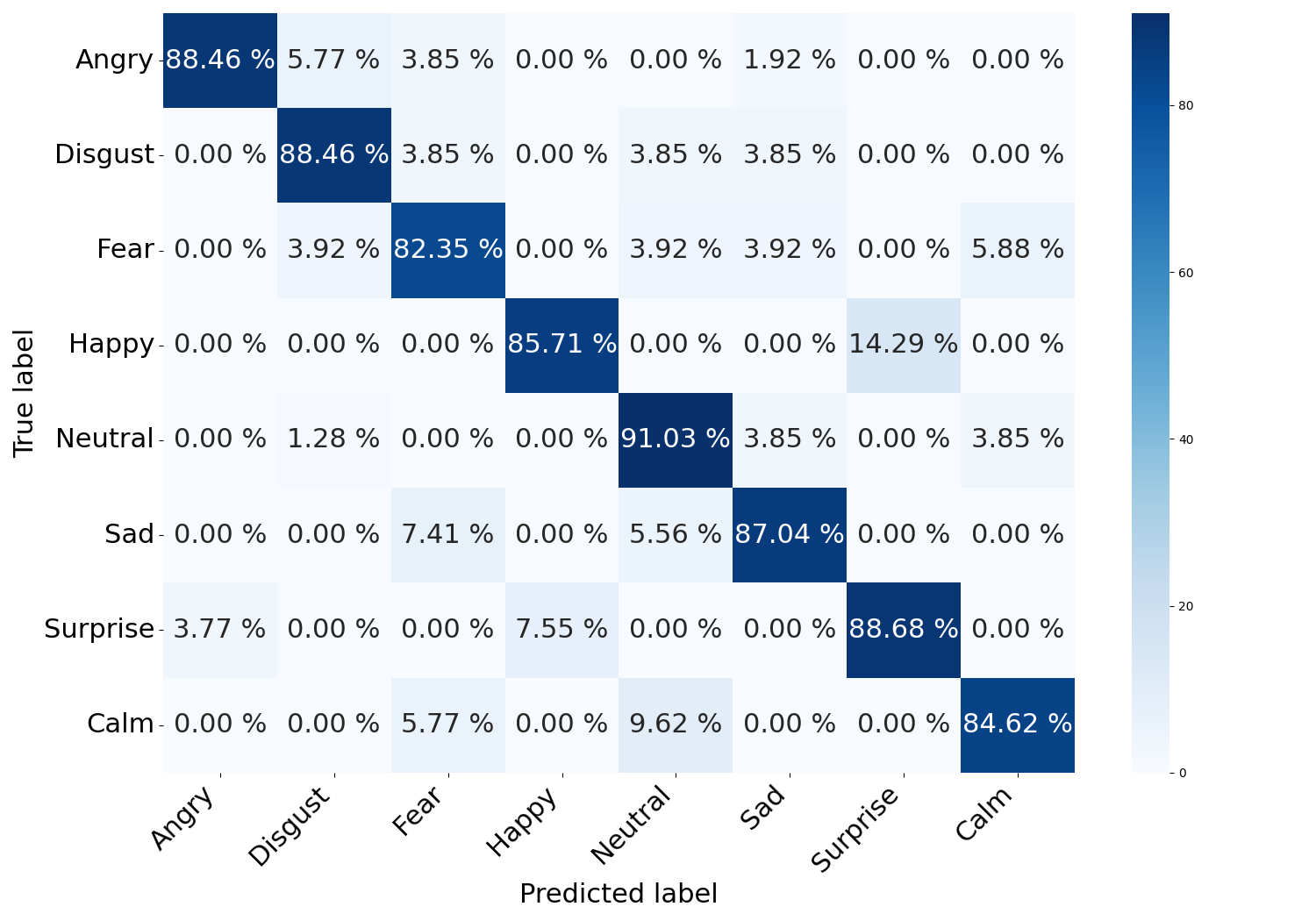}
\caption{Confusion matrix on RAVDESS dataset}\label{fig11}
\end{minipage}\qquad 
\begin{minipage}[b]{.5\textwidth}
\includegraphics[width=\textwidth]{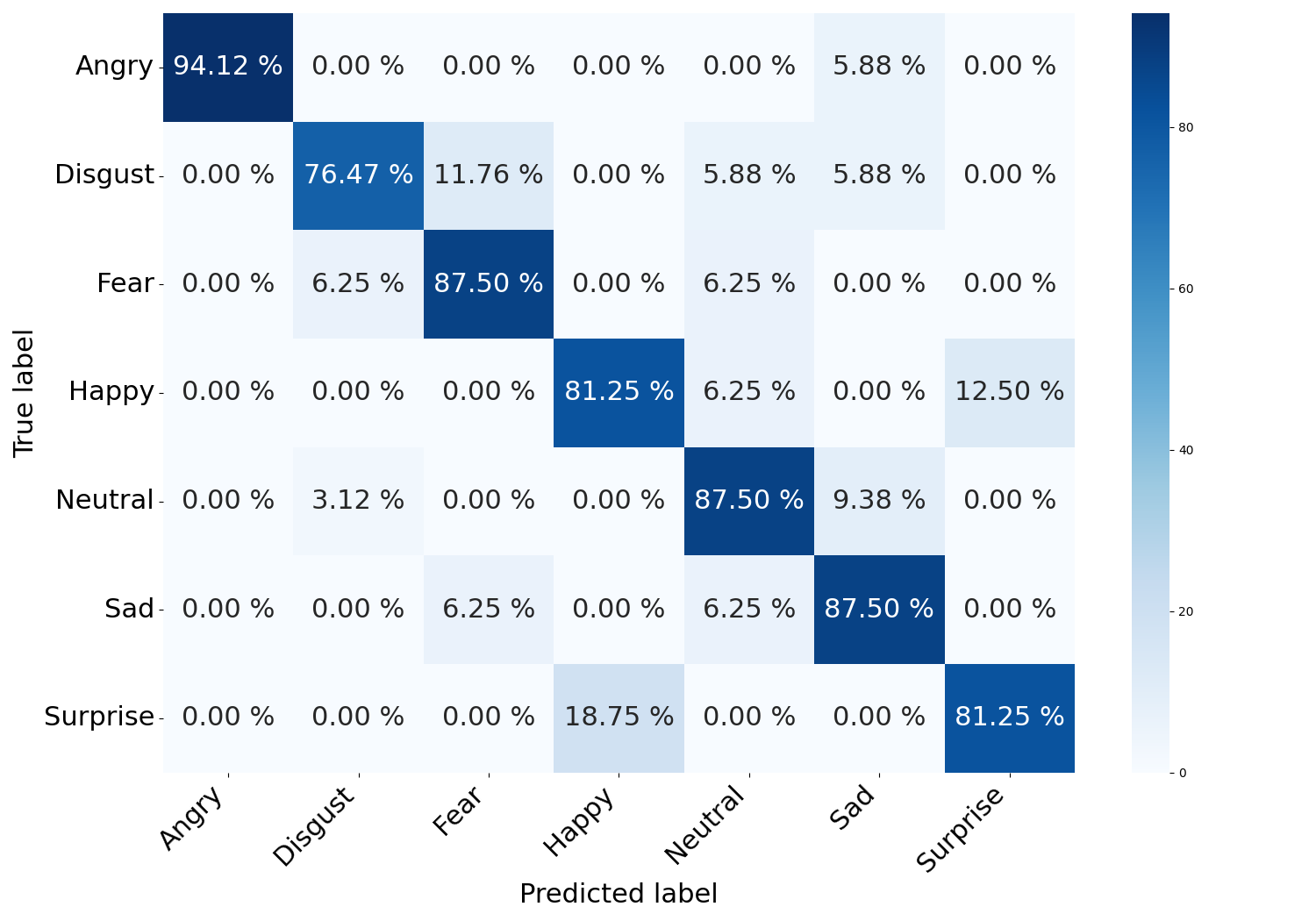}
\caption{Confusion matrix on SAVEE dataset}\label{fig12}
\end{minipage}\\
\end{figure}

Upon detailed examination of the confusion matrices for the RAVDESS (Fig. \ref{fig11}) and SAVEE (Fig. \ref{fig12}) datasets, it's clear that the model demonstrates remarkable proficiency in SER. These matrices, crucial for understanding the model's accuracy in classifying distinct emotions, reveal a consistent pattern of high true positive rates across various emotional states, with a notably low frequency of false positives and false negatives. In RAVDESS dataset, which features a wide range of emotions and complex interactive dialogues, the model adeptly identifies and categorizes emotions with precision, as indicated by the densely populated diagonals in the confusion matrices. Similarly, in SAVEE dataset, despite its unique challenges, the model maintains high accuracy levels. The minor misclassifications observed are typically between emotionally similar categories, such as 'Happy' and 'Surprise', which is a common challenge in SER. However, the overall high correct classification rates across these diverse datasets underscore the model's capability to discern and interpret nuanced emotional expressions in speech. This consistent performance not only attests to the model's reliability but also its adaptability to varied emotional datasets, making it a highly effective tool for SER.

\subsubsection{Statistical evaluation}\label{stat_ev}

To rigorously evaluate the performance of our method relative to the best alternative method across several emotional speech datasets as shown in Fig. \ref{fig13}, we employ independent two-sample t-tests. This statistical method is designed to compare the means of two independent groups (in this case, the performance metrics of two different machine learning models) to ascertain if the observed differences are statistically significant. The t-statistic quantifies the difference between the mean performance metrics relative to the sample variability, providing a basis for assessing the likelihood that such differences arose under the null hypothesis (i.e., no true difference in means). The p-value, derived from the t-statistic and degrees of freedom, offers a measure of the probability of observing the data (or more extreme) if the null hypothesis were true. A conventional alpha level of (\(\alpha = 0.05\)) is used to assess statistical significance, where p-values below this threshold indicate strong evidence against the null hypothesis, suggesting a significant difference in performance metrics. The summarized results, as presented in the statistical evaluation Table \ref{table10}, elucidate the comparative efficacy of our method against the best alternative method with respect to accuracy and F1-score across the four datasets: TESS, EMO-DB, RAVDESS, and SAVEE.

\begin{figure}[t]
\centerline{\includegraphics[width=1\textwidth]{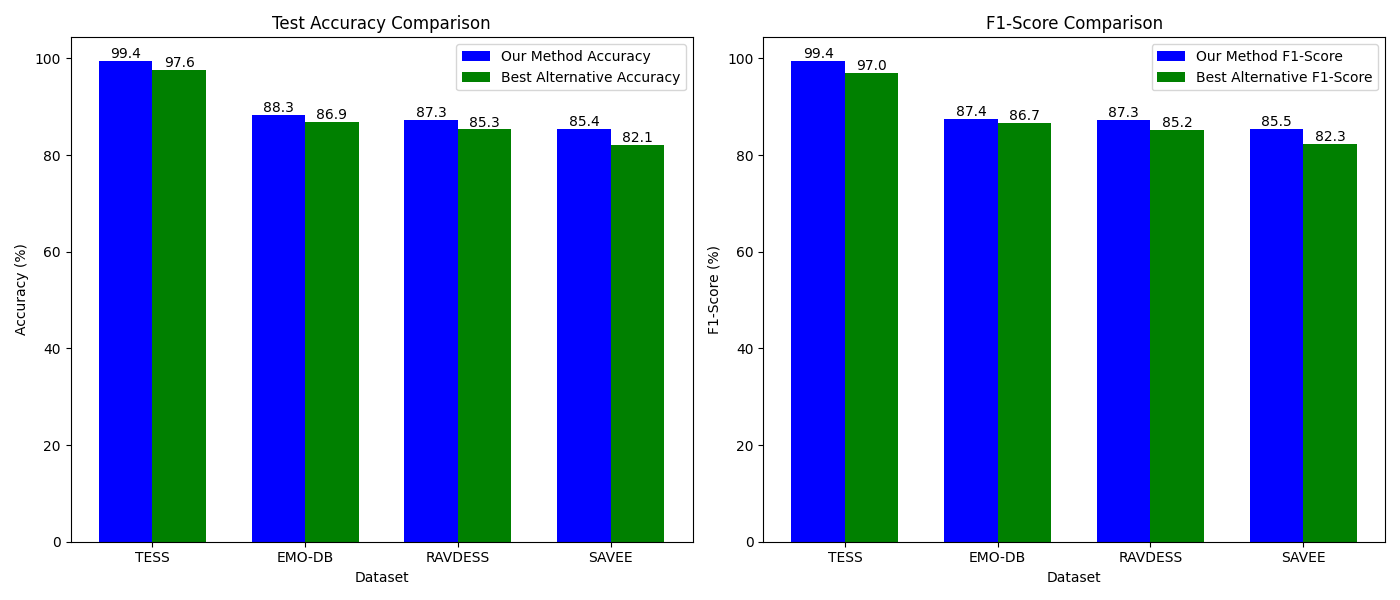}}
\caption{"Our Method" vs. "Best Alternative Method" accuracy and F1-score comparison for all datasets}
\label{fig13}
\end{figure}

\begin{table}[t]
\centering
\begin{tabular}{lcccc}
\hline
\textbf{Dataset} & \multicolumn{2}{c}{\textbf{Accuracy}} & \multicolumn{2}{c}{\textbf{F1-score}} \\
 & \textbf{T-Statistic} & \textbf{P-Value} & \textbf{T-Statistic} & \textbf{P-Value} \\
\hline
\textbf{TESS} & 6.62 & 5.42e-06 & 8.82 & 1.32e-07 \\
\textbf{EMO-DB} & 5.11 & 0.0003 & 2.59 & 0.0244 \\
\textbf{RAVDESS} & 3.55 & 0.0024 & 3.73 & 0.0017 \\
\textbf{SAVEE} & 7.76 & 4.08e-07 & 7.52 & 6.25e-07 \\
\hline
\end{tabular}
\caption{Statistical evaluation of "Our Method" vs. "Best Alternative Method" across different datasets using t-tests. The table presents the t-statistic and p-value for comparisons of both accuracy and F1-score, indicating the statistical significance of the observed differences.}
\label{table10}
\end{table}

The TESS dataset exhibits pronounced differences in both accuracy and F1-score, with t-statistics of 6.62 and 8.82, respectively, and corresponding p-values significantly below the 0.05 threshold. This strongly suggests that our method not only surpasses the best alternative method in terms of overall accuracy but also maintains a superior balance between precision and recall, as indicated by the F1-score. The test on the EMO-DB dataset demonstrates a notable improvement in accuracy (t-statistic: 5.11; p-value: 0.0003) and a significant difference in F1-score (t-statistic: 2.59; p-value: 0.0244), underscoring the effectiveness of our method in recognizing emotional cues within speech with greater reliability and balance. On the RAVDESS dataset, the test shows statistically significant enhancements for both evaluated metrics (Accuracy t-statistic: 3.55, F1-score t-statistic: 3.73), with p-values indicating that these improvements are unlikely to be due to chance. This reinforces the consistency of our method's performance across varied emotional speech contexts. For the SAVEE dataset, the results indicate substantial improvements in both accuracy and F1-score (t-statistics: 7.76 and 7.52, respectively), with exceedingly low p-values, highlighting the robustness of our method in processing and classifying emotional speech with high accuracy and balanced precision-recall performance.

The statistical analysis robustly supports the conclusion that our method significantly outperforms the best alternative method across all evaluated emotional speech datasets. The consistent observation of statistically significant differences in both accuracy and F1-scores, as validated by the t-tests, provides compelling evidence of our method's superior performance. This analysis underscores the effectiveness of our approach for diverse speech emotion recognition tasks.

\section{Discussion}
This work addresses the task of SER and extracting the most relevant features for accurately detecting emotions in speech, a challenge that lacks consensus in the existing literature. The significance of our research lies in bridging this knowledge gap and providing valuable insights into feature selection for SER. We contribute to the field by exploring and identifying features that play a vital role in detecting and distinguishing emotional states in speech, while also emphasizing their interpretability.

Our analysis focuses on identifying features with high discriminative power and informativeness for differentiating between emotional categories. Through a rigorous feature selection process, we aim to identify the most relevant features for SER. Our study highlights the significance of key features, such as MFCCs, which effectively capture the spectral characteristics of speech and have been widely used in speech analysis tasks. Additionally, pitch or fundamental frequency (F0) features emerge as valuable for SER, as variations in pitch convey important emotional cues. Analyzing pitch-related features, such as pitch contour, range, and dynamics, provides valuable information for emotion classification. We also find that energy and intensity measures play a significant role in capturing emotional intensity and arousal, reflecting the overall energy distribution and loudness of speech. Temporal features, including speech rate and duration, demonstrate relevance in capturing temporal patterns and dynamics of emotional speech.

It is important to note that we are aware of our study's limitation, which is its testing solely on acted datasets. While the results obtained are promising, validating our approach on real-world scenarios is crucial for generalizability. Such validation would provide a comprehensive assessment of the effectiveness of our feature boosting approach in different contexts and with various speech samples, uncovering any dataset-specific biases or limitations. Real-world scenarios present additional challenges, including varying recording conditions, speaker characteristics, and noise levels, which can impact the performance of the SER system and the relevance of selected features.

However, our research contributes to the development of a standardized feature set for SER, such as the ones discussed earlier, by presenting a comprehensive analysis of the feature selection process and highlighting the rationale behind specific feature choices. This standardized feature set serves as a foundation for future research in the field, enabling researchers to focus on these key features when designing and implementing robust SER systems. Ultimately, this standardized feature set enhances the accuracy and effectiveness of trustworthy emotion detection in real-world applications.

\section{Conclusion}

This study introduces a novel supervised Speech Emotion Recognition (SER) method based on iterative boosting of designed voice features and their statistical characteristics. The incorporation of feature boosting and explainability is emphasized as crucial for improving the accuracy of SER systems. The proposed method comprises three main modules: the feature boosting module, classification module, and explainability module. Notably, this study uniquely integrates an iterative mechanism based on Explainable Artificial Intelligence (XAI) into the SER framework, which enhances and guides the feature selection process and promotes system interpretability. The presented comprehensive approach strives to strike a balance between leveraging advanced machine learning techniques and addressing the need for transparency and comprehensibility. Experimental results on TESS, EMO-DB, RAVDESS, and SAVEE datasets demonstrate the superiority of the proposed method over state-of-the-art SER methods. Furthermore, the performance surpasses human-level performance (HLP) on the TESS dataset, further validating the significance of the proposed approach. 

In conclusion, this study offers a comprehensive and effective SER approach that highlights the importance of feature boosting and explainability. The method outperforms existing approaches and showcases the potential of incorporating XAI techniques into SER frameworks. Future research directions involve exploring feature boosting within deep learning frameworks, as well as generalizing and evaluating the proposed approach for other classification problems with high dimensionality and feature relevance challenges.

\section{Declarations}
\begin{itemize}
    \item \textbf{Funding:} The authors did not receive support from any organization for the submitted work.
    \item \textbf{Competing interests:} All authors certify that they have no affiliations with or involvement in any organization or entity with any financial interest or non-financial interest in the subject matter or materials discussed in this manuscript.
    \item \textbf{Ethics approval:} Not applicable.
    \item \textbf{Consent to participate:} Not applicable.
    \item \textbf{Consent for publication:} Not applicable.
    \item \textbf{Authors’ contribution:} Alaa Nfissi: Data curation, Formal analysis, Methodology, Software, Validation, Writing – original draft.
    Wassim Bouachir, Nizar Bouguila and Brain L. Mishara: Conceptualization, Methodology, Resources, Supervision, Writing – review \& editing.
    \item \textbf{Data availability and access:} All datasets are publicly available. The EMO-DB is available at \cite{burkhardt2005database}, the TESS is available at \cite{dupuis2010toronto}, the RAVDESS dataset is available at \cite{livingstone2018ryerson}, and the SAVEE dataset is available at \cite{jackson2014surrey}.
    \item \textbf{Code availability:} The source code of this paper is publicly available via this \href{https://github.com/alaaNfissi/Unveiling-the-Hidden-Factors-Explainable-AI-for-Feature-Boosting-in-Speech-Emotion-Recognition}{https://github.com/alaaNfissi/Unveiling-Hidden-Factors-Explainable-AI-for-Feature-Boosting-in-Speech-Emotion-Recognition}.
\end{itemize}


\bibliography{sn-bibliography}

\end{document}